\documentclass[reprint,
nofootinbib,
 amsmath,amssymb,
 aps,
]{revtex4-2}

\usepackage{graphicx}
\usepackage{dcolumn}
\usepackage{bm}
\usepackage{ulem}
\usepackage{braket}
\usepackage{multirow}
\usepackage[dvipsnames]{xcolor}

\begin{document}
\newcommand{\vP}{\mathbf{P}}
\newcommand{\vers}{\mathbf{r''}}
\newcommand{\vRs}{\mathbf{R''}}
\newcommand{\verp}{\mathbf{r'}}
\newcommand{\vRp}{\mathbf{R'}}
\newcommand{\Wcm}{\;\mathrm{W/cm}^2}
\newcommand{\eV}{\;\mathrm{eV}}
\newcommand{\Rep}{\mathrm{Re}\,}
\newcommand{\Imp}{\mathrm{Im}\,}
\newcommand{\vk}{{\mathbf{k}}}
\newcommand{\vkst}{\mathbf{k}_\textrm{st}}
\newcommand{\vqst}{\mathbf{q}_\textrm{st}}
\newcommand{\vi}{\hat{\mathbf{i}}}
\newcommand{\vj}{\hat{\mathbf{j}}}
\newcommand{\vS}{{\mathbf S}}
\newcommand{\vH}{\mathbf{H}}
\newcommand{\vv}{\mathbf{v}}
\newcommand{\ve}{\hat{\mathbf{e}}}
\newcommand{\0}{\mathbf{0}}
\newcommand{\vE}{\mathbf{E}}
\newcommand{\vA}{\mathbf{A}}
\newcommand{\ver}{\mathbf{r}}
\newcommand{\vd}{\mathbf{d}}
\newcommand{\va}{\mathbf{a}}
\newcommand{\vD}{\mathbf{D}}
\newcommand{\vp}{\mathbf{p}}
\newcommand{\vR}{\mathbf{R}}
\newcommand{\vq}{\mathbf{q}}
\newcommand{\vrho}{\mbox{\boldmath{$\rho$}}}
\newcommand{\del}{\mbox{\boldmath{$\nabla$}}}
\newcommand{\valpha}{\mbox{\boldmath{$\alpha$}}}
\newcommand{\vRR}{\{\vR\}}
\newcommand{\vRRp}{\{\vRp\}}
\newcommand{\Ip}{I_\mathrm{p}}
\newcommand{\Up}{U_\mathrm{p}}
\newcommand{\calT}{\mathbf{\cal T}}
\newcommand{\calF}{\mathbf{\cal F}}
\newcommand{\calU}{\mathbf{\cal U}}
\newcommand{\et}{\tilde{e}}
\newcommand{\cm}{\mathrm{c.m.}}
\newcommand{\vro}{\mathbf{\rho}}
\newcommand{\tos}{t_{0s}}
\newcommand{\ts}{t_{s}}
\newcommand{\Ep}{E_{\mathbf{p}}}

\preprint{APS/123-QED}

\title{Below threshold nonsequential double ionization with linearly polarized two-color fields I: symmetry and dominance }
\author{S. Hashim$^1$, D. Habibovi\'{c}$^{2}$, and C. Figueira de Morisson Faria$^{1,3}$}
\affiliation{$^1$Department of Physics and Astronomy, University College London, Gower Street, London, WC1E 6BT, UK\\$^2$
University of Sarajevo, Faculty of Science, Zmaja od Bosne 35, 71000 Sarajevo, Bosnia and Herzegovina\\$^3$ICFO-Institut de Ciencies Fotoniques, The Barcelona Institute of Science and Technology, Av. Carl Friedrich Gauss 3, 08860 Castelldefels, Barcelona, Spain.}

\date{\today}

\begin{abstract}
We investigate laser-induced nonsequential double ionization with linearly polarized bichromatic fields, focusing on the recollision-excitation with subsequent ionization (RESI) mechanism. Using the strong-field approximation, we assess how the symmetries of the field influence the dominant events. Furthermore, we show that, by manipulating the field parameters such as the field frequencies and relative phase between the two driving waves, one can influence the correlated electron-momentum distributions. Specific features of a linearly polarized bichromatic field are that the momentum distributions of the second electron are no longer centered around vanishing momenta and that there may be more than one ionization event per half cycle.  This can be used to confine the RESI distributions to specific momentum regions and to determine a hierarchy of parameters that make an event dominant. 
\end{abstract}

\maketitle


\section{\label{sec:intro}Introduction}

Tailored fields are powerful tools to control laser-induced processes, such as high-order harmonic generation (HHG), above-threshold ionization (ATI), and nonsequential double ionization (NSDI) (for reviews, see, e.g.,  \cite{Brabec2000,Ehlotzky2001,Milos2006}).
Well-known applications, among others, are the \textit{in situ} characterization of attosecond pulses \cite{Dudovich2006,Doumy2009,Dahl2009}, the measurement of tunneling times employing elliptically polarized \cite{Eckle2008,Pfeiffer2011,Pfeiffer2012,Li2013,Ivanov2014,Landsman2013,Torlina2015} and bicircular \cite{Han2018,Eicke2019} fields, temporal gates \cite{Shafir2012,Zhao2013,Henkel2015} or interferometric schemes \cite{Pedatzur2015,Klaiber2018}, the investigation of chiral systems \cite{Smirnova2015JPhysB,Ayuso2018,Ayuso2018II,Baykusheva2018,Habibovic2024},  the study of how the orbital angular momentum (OAM) influences photoelectron vortices  \cite{NgokoDjiokap2015,Bayer2020,Kang2021,Maxwell2021}, 
and the phase-of-the phase spectroscopy using collinear \cite{Skruszewicz2015,Almajid2017,Wuerzler2020} or circularly polarized \cite{Tulsky2018,Tulsky2020} two-color fields.  The fields explored also exhibit a myriad of shapes, including linearly polarized bichromatic fields \cite{Ehlotzky2001}, few-cycle pulses \cite{Brabec2000}, elliptically polarized fields \cite{Landsman2014}, orthogonally polarized two-color (OTC)  \cite{Zhao2013,Zhang2014,Richter2015,Das2013,Das2015,Henkel2015,Li2016,Han2017,Gong2017,Xie2017,Habibovic2021} and bicircular \cite{Milos2000,Smirnova2015JPhysB,Milosevic2016,Mancuso2016,Hoang2017,Eckart2018,Milos2018,Ayuso2018,Ayuso2018II,Baykusheva2018,Eicke2019,Yue2020,Maxwell2021} fields, as well as more exotic shapes such as chiral \cite{Rozen2019} and knotted fields \cite{Pisanty2019}, or even the perfect wave \cite{Chipperfield2009}.

Steering laser-induced processes with shaped fields is enabled by their underlying physical mechanisms, namely laser-induced rescattering or recombination \cite{Corkum1993}. First, an intense laser field considerably distorts the binding potential. This triggers the tunnel ionization of an electron. Once in the continuum, the electron is accelerated by the field, and may or may not be driven back to its parent ion. Direct ATI happens if the electron reaches the detector without further interaction with the core, while rescattered or high-order ATI (HATI) results from its elastic scattering with its parent ion \cite{Becker2018,Becker2002Review,MilosReviewATI}. If, instead, the electron recombines with a target's bound state, it releases its kinetic energy as high-frequency, high-harmonic radiation \cite{Lewenstein1994}. Finally, it may also happen that the electron recollides inelastically with the core, releasing one or more electrons. If a second electron is released, this gives rise to NSDI (for reviews see \cite{Faria2011,Becker2012}). 

A systematic way to understand the imprint of the field and the type of the target on the resulting spectra or photoelectron momentum distributions (PMDs) is provided by symmetry.  Symmetry is a widespread concept in many areas of knowledge, such as chemistry \cite{Cotton1990}, physics \cite{Weyl1952}, and biology \cite{Polak1994}. Not only does it allow us to predict a specific outcome or feature without solving the actual problem, but, in addition, it can be used to derive the selection rules  or explain features that would otherwise remain murky.  In strong-field laser-matter interaction, symmetry has been explored over the past three decades to derive selection rules for HHG  \cite{Alon1998,Milos2015,Liu2016,Neufeld2019,Yue2020} and ATI \cite{Milosevic2016,Busuladvic2017,Habibovic2020}, to determine the shape of photoelectron momentum distributions \cite{Habibovic2021}, and, recently, to explain how different scattering properties of a soft-core and a Coulomb potential manifest themselves in the HATI spectra \cite{Rook2024}. However, the development of a more complete theory, focused on group-theoretical methods and exploring structured light of increasing complexity, as well as other degrees of freedom such as spin, angular momenta is still work in progress (see the perspective articles \cite{Habibovic2024,Neufeld2025}). 

Even linearly polarized fields exhibit temporal symmetries that can be investigated consistently. The best-known of these symmetries is the half-cycle symmetry, which implies that a field is invariant upon a half-cycle translation followed by a reflection about the time axis. In \cite{Rook2022}, we have shown that, further to that, a monochromatic field is also reflection-symmetric about its maxima and crossings. Adding a second wave may break or retain these symmetries, depending on its frequency and relative phase. If the half-cycle symmetry is broken, one of the other two symmetries is automatically broken, while, if it is retained, the other two may either be broken or retained. 
For orthogonally polarized fields, symmetries are often studied by constructing compound systems, using properties of the field and sometimes of the target. The field exhibits temporal and geometric symmetries, which were studied systematically in \cite{Neufeld2019}. Furthermore, when the geometry of the target must be taken into account, it must be considered jointly with the symmetry of the field \cite{Liu2016,Busuladvic2017,Habibovic2020}. 

For laser-induced processes involving more than one active electron, such as NSDI, symmetries are considerably less explored. However, there have been plenty of studies of NSDI in few-cycle pulses \cite{Liu2004,Bergues2012,Huang2016,Kubel2016,Chen2017}, circularly polarized fields \cite{Fu2012,Huang2018}, polarization gated fields \cite{Quan2009}, OTC fields \cite{Zhang2014,Mancuso2016,Song2018}, or few-cycle counter-rotating two-color circularly or elliptically polarized (TCCP, TCEP) laser fields \cite{Pang2020,Ge2023,Liu2024}.  The overwhelming majority of these investigations have been performed using classical-trajectory methods, and have focused on the shape of the electron-momentum distributions, which is determined by, for instance, final-state electron-electron repulsion \cite{Huang2016}, different types of recollisions and pathways \cite{Chen2017},  temporal windows for recollision dynamics \cite{Fu2012,Zhang2014,Mancuso2016}. Similarly, quantum mechanical approaches have also been mainly used to address similar questions. For instance, the full solution of the time-dependent Schr\"odinger equation (TDSE) was used to assess how the shapes of the electron momentum distributions are affected by the type of electron-electron interaction \cite{Lein2000,Parker2006} or the field \cite{Baier2006,Baier2007}. The same holds for early work using the strong-field approximation (SFA) \cite{Faria2004,Faria2004b,Faria2005,Faria2008}, or studies employing the quantitative rescattering theory (QRS) \cite{Chen2010,Chen2019,Chen2021,Chen2022}. 

Nonetheless, group-theoretical arguments are much less explored for correlated momentum distributions in NSDI. Still, some symmetries have been identified.  For instance, if the NSDI process is electron-impact (EI) ionization, which prevails if the first electron returns with enough energy to make the second electron overcome the ionization potential of the singly ionized target, the electron momentum distributions, as functions of the electron momentum components $p_{1\parallel}$ and  $p_{2\parallel}$  parallel to the laser-field polarization, are symmetric concerning reflections about the main diagonal $p_{1\parallel}=p_{2\parallel}$, occupy the first and third quadrant of the parallel momentum plane, and, for half-cycle symmetric fields, are symmetric upon $(p_{1\parallel},p_{2\parallel}) \leftrightarrow (-p_{1\parallel},-p_{2\parallel}) $ \cite{Faria2004,Faria2005}. 
If, on the other hand, the second electron is dislodged by recollision-excitation with subsequent ionization (RESI), in which the second electron is excited by the first and is freed with a time delay, a myriad of shapes has been identified.  These include electron momentum distributions occupying the second and fourth quadrants of the $p_{1\parallel}p_{2\parallel}$ plane, distributions occupying the axes $p_{n\parallel}=0$, $n=1,2$, and/or the diagonals $p_{1\parallel}=\pm p_{2\parallel}$, or distributions concentrated in the positive or negative parallel momentum half-plane. RESI is prevalent in the below-threshold regime, for which the electron's kinetic energy, upon return, is only sufficient to promote the second electron to an excited state.  In particular, quantum-mechanical studies based on the SFA revealed fourfold symmetric RESI distributions, whose shape depends on the geometry of the bound state to which the second electron was excited. The fourfold symmetry is broken if the field is not half-cycle symmetric, such as for the few-cycle pulses \cite{Faria2012,Shaaran2012,Hashim2024}. In this case, the correlated electron momentum distributions will be shaped by the dominant events and the momentum regions they occupy. Additionally, in \cite{Hao2014}, it was shown that fourfold symmetry is broken if quantum interference between different excitation channels is incorporated. These results were confirmed and extended in our previous work, in which we identified various types of quantum interference in the RESI distributions for monochromatic fields \cite{Maxwell2015,Maxwell2016} and few-cycle pulses \cite{Hashim2024}. 
 
In the present work, we investigate RESI with bichromatic driving fields composed of a wave of frequency $\omega$ and its second or third harmonic, using the symmetry arguments from our previous publication \cite{Rook2022}. These fields are widely known as ($\omega$,$2\omega$) and ($\omega$,$3\omega$)  fields, respectively. By changing the relative phase between both waves, the field-specific symmetries may be either broken or retained. In the present publication, we focus on the influence of the symmetries and dominant events on the PMDs and perform incoherent sums. In particular, we assess how the symmetries dictated by the field are mapped into the RESI electron momentum distributions, as functions of the electron-momentum component parallel to the driving-field polarization.  Quantum interference is expected to break some field symmetries, and would mask the these features \cite{Hao2014,Maxwell2015,Maxwell2016}. In a complementary paper, we show that it leads to a myriad of fringe shapes, whose interpretation is beyond the scope of the present work \cite{hashim2024c}.  

Throughout, we use the SFA, employing the transition amplitude derived in \cite{Shaaran2010} for RESI, in which we incorporated electron-electron correlation and excitation. Although the SFA relies on several simplifications, such as neglecting the residual binding potentials in the electrons' continuum propagation, it provides a good testing ground for the features we intend to study. First, the SFA allows us to single out the specific scattering process leading to RESI, while different physical mechanisms are difficult to disentangle in \textit{ab initio} methods such as the full solution of the time-dependent Schr\"odinger equation (see the perspective article \cite{Armstrong2021} for the advantages and shortcomings of numerical and analytical methods). Second, because the SFA is a Born-type approach, it provides a clear-cut definition of direct and rescattered processes. These definitions become blurred in Coulomb-distorted approaches, for which there are hybrid quantum pathways that do not fit in either category \cite{Yan2010,Lai2015a,Maxwell2017,Maxwell2018,Bray2021}.  Third, if the transition amplitude is calculated using the steepest descent method, specific quantum pathways may be associated with electron orbits, which provides a great deal of physical insight. Fourth, due to being semi-analytic, the SFA exhibits features that can be switched on an off at will. Depending on the context, making it deliverately less accurate sheds light on important physics. Finally, in the SFA framework, RESI can be viewed as two time-ordered ATI-like processes. This is a physical picture that is useful for studying symmetries and has been made more obvious by the SFA, as it entails precise definitions of scattering. 

This article is organized as follows. In Sec.~\ref{sec:backgd}, we bring the necessary background to understand the subsequent results. This includes the SFA transition amplitude for RESI and the saddle-point equations, the three symmetries exhibited by the linearly polarized monochromatic field, and which of those are broken or retained if a two-color field is considered. 
Sec.~\ref{sec:dominance} is devoted to determining the dominant events, and linking them to the existing field symmetries. Subsequently, in Sec.~\ref{section:momdists}, we assess how these findings fit together in correlated electron momentum distributions. Finally, in Sec.~\ref{sec:conclusions}, we summarize this work and state our conclusions.
\section{Background}
\label{sec:backgd}
\subsection{Transition amplitude}
The SFA  transition amplitude for RESI and an arbitrary excitation channel $\mathcal{C}$ reads
\begin{eqnarray}
&&M^{(\mathcal{C})}(\mathbf{p}_{1},\mathbf{p}_{2})=\hspace*{-0.2cm}\int_{-\infty }^{\infty
}dt\int_{-\infty }^{t}dt^{^{\prime }}\int_{-\infty }^{t^{\prime
}}dt^{^{\prime \prime }}\int d^{3}k  \notag \\
&&\times V^{(\mathcal{C})}_{\mathbf{p}_{2}e}V^{(\mathcal{C})}_{\mathbf{p}_{1}e,\mathbf{k}g}V^{(\mathcal{C})}_{\mathbf{k}%
	g}\exp [iS^{(\mathcal{C})}(\mathbf{p}_{1},\mathbf{p}_{2},\mathbf{k},t,t^{\prime },t^{\prime
	\prime })],  \label{eq:Mp}
\end{eqnarray}
where
\begin{eqnarray}
&&S^{(\mathcal{C})}(\mathbf{p}_{1},\mathbf{p}_{2},\mathbf{k},t,t^{\prime },t^{\prime \prime
})=  \notag \\
&&\quad E^{(\mathcal{C})}_{\mathrm{1g}}t^{\prime \prime }+E^{(\mathcal{C})}_{\mathrm{2g}}t^{\prime
}+E^{(\mathcal{C})}_{\mathrm{2e}}(t-t^{\prime })-\int_{t^{\prime \prime }}^{t^{\prime }}%
\hspace{-0.1cm}\frac{[\mathbf{k}+\mathbf{A}(\tau )]^{2}}{2}d\tau  \notag \\
&&\quad -\int_{t^{\prime }}^{\infty }\hspace{-0.1cm}\frac{[\mathbf{p}_{1}+%
	\mathbf{A}(\tau )]^{2}}{2}d\tau -\int_{t}^{\infty }\hspace{-0.1cm}\frac{[%
	\mathbf{p}_{2}+\mathbf{A}(\tau )]^{2}}{2}d\tau  \label{eq:singlecS}
\end{eqnarray} 
is the semiclassical action. 
Equations~\eqref{eq:Mp} and \eqref{eq:singlecS} have been derived in detail in \cite{Shaaran2010,Shaaran2010a} and 
correspond to a process in which an electron, initially bound in a state of energy $-E^{(\mathcal{C})}_{1g}$, is freed in the continuum at a time $t^{\prime\prime}$. Subsequently, at a time $t^{\prime}$, it returns to its parent ion with intermediate momentum $\mathbf{k}$ and excites a second electron from a bound state with energy  $-E^{(\mathcal{C})}_{2g}$ to a state with energy  $-E^{(\mathcal{C})}_{2e}$. The first electron then leaves, reaching the detector with the final momentum $\mathbf{p}_1$. The second electron is freed at a later time $t$, and has final momentum $\mathbf{p}_2$. 

In the SFA, all information about the target geometry and the interactions is embedded in the 
 prefactors  $V^{(\mathcal{C})}_{\mathbf{k}g}$, $V^{(\mathcal{C})}_{\mathbf{p}_1e,\mathbf{k}g}$ and $V^{(\mathcal{C})}_{\mathbf{p}_2e}$. 
The prefactor
     \begin{eqnarray}\label{eq:pre3}
V^{(\mathcal{C})}_{\mathbf{k}g} &=& \bra{\mathbf{k}+\mathbf{A}(t'')}V\ket{\psi_{1g}^{(\mathcal{C})}}\\ &= &\frac{1}{(2\pi)^{3/2}}\int d^3r_1e^{-i[\mathbf{k} +\mathbf{A}(t'')]\cdot\mathbf{r}_1}V(\mathbf{r}_1)\psi_{1g}^{(\mathcal{C})}(\mathbf{r}_1), \notag
\end{eqnarray}
where $V(\mathbf{r}_1)$ is the neutral atom's binding potential, and $\psi_{1g}^{(\mathcal{C})}$ is the ground-state wave function for the first electron,  
is associated with the ionization of the first electron.  This electron, initially in $\ket{\psi_{1g}^{(\mathcal{C})}}$, is released in an intermediate Volkov state $\ket{\mathbf{k}+\mathbf{A}(t'')}$. 

The prefactor 
\begin{eqnarray}
\label{eq:Vp1ekg}V^{(\mathcal{C})}_{\mathbf{p}_1e,\mathbf{k}g}\hspace*{-0.1cm}&=& \hspace*{-0.1cm} \bra{\mathbf{p}_1,\psi_{2e}^{(\mathcal{C})}}V_{12}
\ket{\mathbf{k},\psi_{2g}^{(\mathcal{C})}}
    \\
    &=&\hspace*{-0.1cm} \frac{V_{12}(\mathbf{p}_1-\mathbf{k})}{(2\pi)^{3/2}}\hspace*{-0.1cm}\int \hspace*{-0.1cm}d^3r_2e^{-i(\mathbf{p}_1-\mathbf{k})\cdot \mathbf{r}_2}\psi_{2e}^{*(\mathcal{C})}(\mathbf{r}_2)\psi_{2g}^{(\mathcal{C})}(\mathbf{r}_2),  \notag 
\end{eqnarray}
where
\begin{equation}
    V_{12}(\mathbf{p}_1 - \mathbf{k}) =  \frac{1}{(2\pi)^{3/2}}\int d^3rV_{12}(\mathbf{r})\exp[-i(\mathbf{p}_1-\mathbf{k})\cdot\mathbf{r}]
\end{equation}
is the electron-electron interaction in momentum space, $\mathbf{r} = \mathbf{r}_1-\mathbf{r}_2$, and $V_{12}(\mathbf{r})$, taken to be of contact type, describes the interaction by which the second electron is excited.  The wave functions $\braket{\mathbf{r}_2|\psi_{2e}^{(\mathcal{C})}}=\psi_{2e}^{(\mathcal{C})}(\mathbf{r}_2)$ and $\braket{\mathbf{r}_2|\psi_{2g}^{(\mathcal{C})}}=\psi_{2g}^{(\mathcal{C})}(\mathbf{r}_2)$ are associated with the excited and ground states of the second electron, respectively. Finally, the prefactor
\begin{eqnarray}
    V^{(\mathcal{C})}_{\mathbf{p}_2e} &=& \bra{\mathbf{p}_2+\mathbf{A}(t)}V_{\mathrm{ion}}\ket{\psi_{2e}^{(\mathcal{C})}}      \label{eq:Vp2e}\\ &=& \frac{1}{(2\pi)^{3/2}}\int d^3r_2V_{\mathrm{ion}}(\mathbf{r}_2)e^{-i[\mathbf{p_2}+\mathbf{A}(t)]\cdot\mathbf{r}_2}\psi_{2e}^{(\mathcal{C})}(\mathbf{r}_2),\notag
\end{eqnarray}
where $V_{\mathrm{ion}}(\mathbf{r}_2)$ is the potential of the singly ionized target, describes the ionization of the second electron. One should note that Eqs.~\eqref{eq:pre3} and \eqref{eq:Vp2e} are written in the length gauge. In their velocity-gauge counterparts, the vector potentials are removed from the final states due to the unitary transformation from the length to the velocity gauge, which is a translation in momentum space. This means that $\ket{\psi^{(V)}_{\mathbf{k}}}=\ket{\mathbf{k}}$ instead of $\ket{\psi^{(L)}_{\mathbf{k}}}=\ket{\mathbf{k}+\mathbf{A}(t'')}$ and $\ket{\psi^{(V)}_{\mathbf{p}_2}}=\ket{\mathbf{p}_2}$ instead of $\ket{\psi^{(L)}_{\mathbf{p}_2}}=\ket{\mathbf{p}_2+\mathbf{A}(t)}$. Nonetheless, ionization occurs most probably around a field maxima, which, for monochromatic fields and few-cycle pulses, implies that $|\mathbf{A}(\tau)|\ll 1$, with $\tau=t'',t$. From the practical point of view, this means that the length- and velocity-gauge prefactors lead to very similar results, although strictly speaking the SFA is not gauge invariant. Thus, it is a reasonable approximation to neglect the vector potential in Eqs. \eqref{eq:pre3} and \eqref{eq:Vp2e}. This approximation has been discussed in detail in \cite{Shaaran2010} and has been used in our previous publications. 
However, in some instances, it may break down for the two-color driving fields. 

In those cases, it will be necessary to incorporate the vector potential in the length-gauge prefactors. This will render the prefactors calculated using hydrogenic bound-state wave functions singular due to the saddle-point equation \eqref{eq:spati} given below. The explicit expression for these prefactors are given in  \cite{Shaaran2010,Maxwell2015,Hashim2024}. 
This problem may be overcome by either exponentializing the prefactors to eliminate the singularity and incorporating it in the action as a logarithmic term \cite{Faria2005}, or by approximating the hydrogenic wave functions using a Gaussian basis set. The latter method has been employed in \cite{Augstein2010,Shaaran2011} in the context of diatomic molecules, and will be used in Sec.~\ref{sec:prefactors} whenever necessary. In case the singularity is absent, we will employ the prefactors calculated in \cite{Shaaran2010,Maxwell2015,Hashim2024} for hydrogenic wave functions. The explicit expression for $V_{\mathbf{p}_2e}$ computed using a Gaussian basis set, together with a test showing that the results are essentially the same as if using hydrogenic functions, is provided in the Appendix.

\subsection{Saddle-point method}

The multiple integral that appears in the transition amplitude \eqref{eq:Mp} is solved using the saddle-point method \cite{Milosevic2024TopRev}, which requires finding the values of the integrating variables such that the action is stationary. This leads to $\partial S(\vp_1,\vp_2,\vk;t,t',t'')/\partial t=\partial S(\vp_1,\vp_2,\vk;t,t',t'')/\partial t'=\partial S(\vp_1,\vp_2,\vk;t,t',t'')/\partial t''=0$ and $\partial S(\vp_1,\vp_2,\vk;t,t',t'')/\partial \vk=\textbf{0}$, which give the saddle-point equations 
\begin{equation}\label{sphati1}
[\vk+\vA(t'')]^2=-2E_{1g},
\end{equation}  
\begin{equation}\label{sphati2}
\vk=-\frac{1}{t'-t''}\int_{t''}^{t'}d\tau\vA(\tau),
\end{equation}
\begin{equation}\label{sphati3}
[\vp_1+\vA(t')]^2=[\vk+\vA(t')]^2-2(E_{2g}-E_{2e}),
\end{equation}
for the first electron, while for the second electron, the corresponding saddle-point equation is
\begin{equation}\label{eq:spati}
[\vp_2+\vA(t)]^2=-2E_{2e}.
\end {equation}
Equation \eqref{sphati1} represents the energy-conservation condition at time $t''$, while equation  \eqref{sphati2} represents the constraint on the intermediate momentum $\vk$ of the first electron, i.e., it gives the condition which has to be satisfied in order for the electron to return to the parent ion. Moreover, equation \eqref{sphati3}  is the energy-conservation condition at time $t'$. Finally, equation \eqref{eq:spati} describes the tunneling ionization of the second electron at some later time $t$.  Eqs. \eqref{sphati1} and \eqref{eq:spati} have no real solution, which is a consequence of tunneling having no classical counterpart. However, in the limit $E_{1g} \rightarrow 0$, $E_{2e} \rightarrow 0$,  lead to the mappings  $\mathbf{k}=-\mathbf{A}(t'')$, $\mathbf{p}_2=-\mathbf{A}(t)$, widely used in classical-trajectory methods, and real $t''$, $t$. This is the classical limit of the SFA, which allows simplifications that can be used for determining kinematic constraints or for constructing graphical methods to identify approximate ionization and rescattering times \cite{Habibovic2025}. 

If written in terms of the momentum components $p_{n\parallel}$ and $\mathbf{p}_{n\perp}$, $n=1,2$, parallel and perpendicular to the laser-field polarization, saddle-point equations \eqref{sphati3} and \eqref{eq:spati} shed some light on the momentum regions occupied by the correlated electron momentum distributions. 

For a lineary polarized fied, Eq.~\eqref{sphati3} can be re-written as 
\begin{equation}
    [p_{1\parallel}+A(t')]^2=[k+A(t')]^2-[2(E_{2g}-E_{2e})+\vp^2_{1\perp}], \label{eq:sp3ppar}
\end{equation}
which, in terms of the momentum components of the first electron, gives a sphere centered at $(p_{1x},p_{1y},p_{1\parallel})=(0,0,-A(t'))$, where $\mathbf{p}_{1\perp}=p_{1x}\hat{e}_x+p_{1y}\hat{e}_y$, whose radius is real if $[k+A(t')]^2>2(E_{2g}-E_{2e})$, A real radius means that the process is classically allowed. This concept can be used to define a momentum region for which there is a classical counterpart, known as the `classically allowed region'. Within this region, the RESI probability density is appreciable, while outside this region it is exponentially decaying. For details within the framework of RESI and electron-impact ionization see our previous publications \cite{Faria2012,Hashim2024} and \cite{Faria2003,Faria2004b}, respectively.  

For constant $\mathbf{p}_{1\perp}$, the right-hand side of Eq.~\eqref{eq:sp3ppar} shows that it mainly adds a term to the energy gap $E_{2g}-E_{2e}$, effectively decreasing the classically allowed region, so that an upper bound can be obtained for $\mathbf{p}_{1\perp}=\mathbf{0}$. This bound suggests a classically allowed region centered at $p_{1\parallel}=-A(t')$, whose extension is determined by the difference between the maximal kinetic energy of the first electron upon return and the energy gap. 

For the second electron, Eq.~\eqref{eq:spati} reads 
\begin{equation}\label{eq:sp4ppar}
[p_{2\parallel}+A(t)]^2=-2E_{2e}-\mathbf{p}_{2\perp},
\end {equation}
where, similarly, the perpendicular momentum effectively shifts the bound state energy for the second electron if it is kept fixed. Eq.~\eqref{eq:sp4ppar} has no classical counterpart but also describes a sphere centered around the most probable momentum. Therefore, we can infer that the probability density, as a function of $p_{2\parallel}$, is centered around $p_{2\parallel}=-A(t)$. Bringing these constraints together means that the length and width of the correlated two-electron momentum distributions are determined by the first and second electron, respectively. Electron indistinguishability requires symmetrization upon momentum exchange, which means that we must consider $M^{(\mathcal{C})}(\mathbf{p}_{2},\mathbf{p}_{1})$ in addition to $M^{(\mathcal{C})}(\mathbf{p}_{1},\mathbf{p}_{2})$. For details on both constraints for monochromatic fields and few-cycle pulses see \cite{Shaaran2010a} and \cite{Hashim2024}, respectively.

\subsection{Correlated momentum distributions}

Here, we aim at computing the correlated two-electron probability density as a function of the momentum components $p_{n\parallel}$, $n=1,2$ parallel to the driving-field polarization. Its explicit expression reads
\begin{align}
	\mathcal{P}(p_{1\parallel},p_{2\parallel})= \iint d^2 p_{1\perp}d^2 p_{2\perp}\mathcal{P}(\mathbf{p}_{1},\mathbf{p}_{2}), \label{Eq:Channels}
\end{align}
where $\mathcal{P}(\mathbf{p}_{1},\mathbf{p}_{2})$ is the fully resolved two-electron momentum probability density, and the transverse momentum components have been integrated over.  In this paper, we focus on a single excitation channel ${(\mathcal{C})}$ and incoherent sums of probability densities, so that $\mathcal{P}(\mathbf{p}_{1},\mathbf{p}_{2})=\mathcal{P}^{(\mathcal{C})}(\mathbf{p}_{1},\mathbf{p}_{2})$ and 
\begin{equation}
\mathcal{P}^{(\mathcal{C})}_{(\mathrm{ii})}(\mathbf{p}_{1},\mathbf{p}_{2})=\sum_{\varepsilon}\left[\left|M_{\varepsilon}^{(\mathcal{C})}(\mathbf{p}_1,\mathbf{p}_2)\right|^2\hspace*{-0.2cm}+\hspace*{-0.1cm}\left|M_{\varepsilon}^{(\mathcal{C})}(\mathbf{p}_2,\mathbf{p}_1)\right|^2\right], 
\label{eq:1ii}
\end{equation}
where the subscripts $(\mathrm{ii})$ indicate that we are summing incoherently upon events $\varepsilon$ and upon the two contributions due to the electron symmetrization. In the subsequent sections, we will omit both subscripts as we are not studying coherent sums in the present work.  

Furthermore, it is useful to compute partial momentum distributions for each electron, given by 
\begin{equation}
   \mathcal{P}^{(n)}(p_{n\parallel})=\int d^2 p_{n \perp}|M^{(n)}(\mathbf{p}_n)|^2,
   \label{eq:partialdistr}
\end{equation}
with $n=1,2$, and $M^{(n)}(\mathbf{p}_n)$ the amplitude which corresponds to a single electron.
\subsection{Model and field symmetries}
\label{sec:model}
Next, we briefly state the target and the field used in this work.  We consider argon, for which the first ionization potential is $E^{(\mathcal{C})}_{1g}=0.58$ a.u. for all channels. An electron tunnels from the outer shell, so that the target becomes singly ionized. The corresponding electronic configuration is
$ \quad 1s^2\, 2s^2\, 2p^6\, 3s^2\, 3p^5$, so that $E^{(1)}_{2g}= E^{(2)}_{2g}= 1.016$ a.u., which is the second ionization potential. Most of our computations are performed for the $3s \rightarrow 3p$ excitation channel (electron configuration $3s3p^6$), with $E^{(1)}_{2e}=0.52$ a.u., which is the deepest bound state for this specific target. In Sec.~\ref{sec:prefactors}, we also take the $3p \rightarrow 4s$ $(3p^44s)$ excitation pathway, with $E^{(2)}_{2e}=0.40$ a.u., in order to assess the influence of an $s$ excited state in correlated electron momentum distributions. Choosing deeply bound excited states aims at avoiding a too strong influence of the field gradient, which is critical if the excited state is loosely bound \cite{Hashim2024}. 

We use linearly polarized bichromatic fields with commensurate frequencies $r\omega$ and $s\omega$, where $r,s$ are chosen to be co-prime integers. For this field, the vector potential can be written in the form
\begin{equation}
   \mathbf{A}_{r, s, \xi, \phi}(t)= \frac{2 \sqrt{U_p} }{\sqrt{\frac{1}{r^2}+\frac{\xi ^2}{s^2}}}\left[\frac{\xi  \cos (s\omega t+\phi)}{s}+\frac{\cos (r \omega t)}{r}\right]\hat{e}_z,
   \label{eq:Afield}
\end{equation}
where 
\begin{equation}
    U_p=\frac{E^2_{r\omega}} {4\omega^2}\left(\frac{1}{r^2}+\frac{\xi ^2}{s^2}\right)
    \label{eq:epond}
\end{equation}
is the ponderomotive energy and the field is polarized along $\hat{e}_z$. Also, in Eqs.~\eqref{eq:Afield} and \eqref{eq:epond}, $E_{r\omega}$ is the amplitude associated with the wave of frequency $r\omega$, $\phi$ is the relative phase between the two driving waves, and $\xi=E_{s\omega}/E_{r\omega}$ is the field-strength ratio. We assume that our driving field is a long pulse with a flat envelope. 

Depending on the values of these parameters, the field exhibits different symmetries with regard to three main types of transformations and the combinations thereof. These transformations are reflections about the time axis [for which $E(t)$ and $A(t)$ vanish] shifting $E(t) \rightarrow -E(t)$ and $A(t) \rightarrow -A(t)$, denoted by $\mathcal{F}$, translations in time, referred to as $\mathcal{T}_T(\tau_T)$, and reflections about specific times, called $\mathcal{T}_R\left(\tau_{R}\right)$, where the arguments $\tau_T$ and $\tau_{R} $ give the time interval considered in the translation and the time about which the reflection is performed, respectively. 

For instance, a monochromatic linearly polarized field of frequency $r\omega$ exhibits three symmetries. First, it is symmetric with regard to a translation by half a cycle followed by a reflection about the time axis. Summarizing, $\mathcal{F}\mathcal{T}_T\left(\frac{T}{2}\right)E(t)=E(t)$. This is known as the half-cycle symmetry and is usually written as $E(t\pm T/2)=-E(t)$. Second, it is symmetric regarding a time reflection around its extrema, so that $\mathcal{T}_R\left(\tau_{ex}\right)E(t)=E(t)$, where $\tau_{ex}$ are the times for which the extrema occur. Finally, the field remains invariant with respect to a time reflection around its zero crossings followed by a reflection with regard to the time axis. This implies that $\mathcal{F}\mathcal{T}_R\left(\tau_{cr}\right)E(t)=E(t)$, where $\tau_{cr}$ are the times for which the field zero crossings happen. 

Adding a collinearly polarized second wave of frequency $s\omega$ may retain these symmetries or break some of them.  
If $r+s$ is even, for instance, for the $(\omega,3\omega)$ field, the half-cycle symmetry is retained, while, depending on the relative phase, the other symmetries may be retained or broken. If $r+s$ is odd, for example, for the $(\omega,2\omega)$ field, the half-cycle symmetry \textit{and} one of the other two symmetries is broken. For details see our previous publication \cite{Rook2022}. 

\begin{figure*}[h!tb]
\hspace*{-0.7cm}
\includegraphics[width=1.1\textwidth]{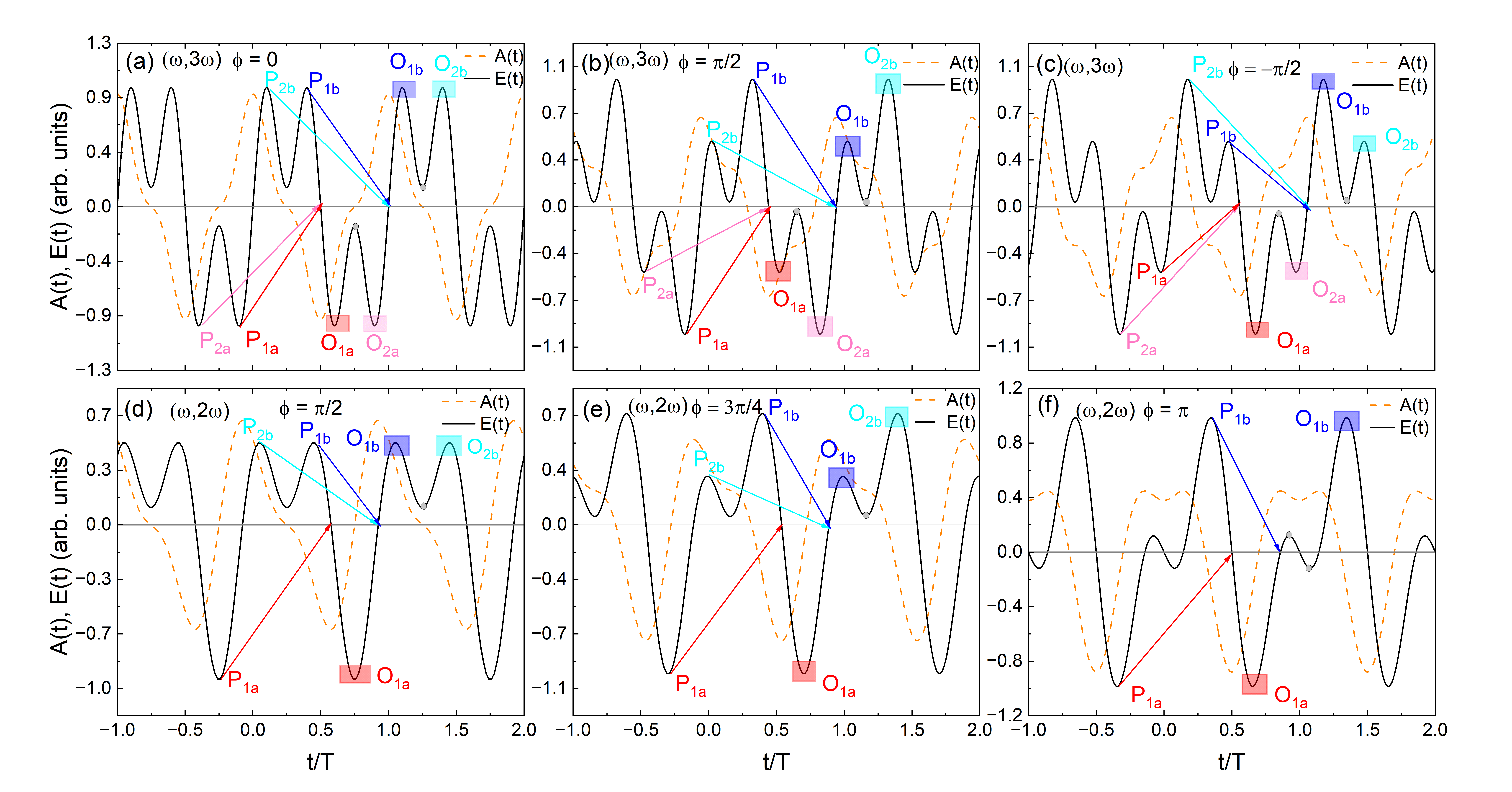} 
    \caption{Electric field (black solid line) and the corresponding vector potential (orange dashed line) as functions of time for the ($\omega$,$3\omega$) [panels (a), (b), and (c)], and ($\omega$,$2\omega$) [panels (d), (e), and (f)] bichromatic linearly polarized field. The ratio of the field amplitudes is $\xi=0.8$, and the relative phase is indicated in the panels.  The approximate values of the real part of the ionization and rescattering times of the first electron associated with the saddle-point pairs which lead to the most significant contributions to the photoelectron yield are indicated by the arrows, while the shaded rectangles correspond to the approximate values of the real part of the ionization time of the second electron. The subscript $n=1,2$ in the pairs $P_{n\mu}$ classifies them in increasing order of excursion times in the continuum, i.e., the excursion time for the pairs  $P_{1\mu}$ are smaller than those for pairs  $P_{2\mu}$. The index $\mu=a,b$ refers to the first and second half cycle taken into consideration, respectively. The red and pink, and the blue and cyan arrows indicate rescattering events populating the positive and negative momentum regions, respectively. The colors of the shaded rectangles match those of the arrow, but, rather, indicate that the orbits $O_{jk}$ of the second electron are associated with a specific pair for the first electron, instead of referring to the momentum region they populate. The subscript $j=1,2$ refers to how close the event is regarding the time of rescattering of the first electron, and the indices $a$ and $b$ refer to the first and second half cycle considered, respectively. The gray dots indicate irrelevant ionization events. The electric fields have been normalized to their maximum amplitude in each panel. }
    \label{fig:fields}
\end{figure*}

\section{Symmetries and dominant events}\label{sec:dominance}

Next, we investigate how the specific field shape considered in this paper, corresponding to the vector potential given by Eq.~\eqref{eq:Afield}, influences the RESI momentum distributions. To achieve this goal, it is necessary to identify the dominant ionization and rescattering events for the first electron, followed by their counterparts for the second electron's ionization. These events are first mapped by employing classical arguments and inspecting the field.  This can be done as, in the limit of vanishing binding energies, the saddle-point equations give the classical times of an electron in the presence of the driving field (see, e.g., \cite{Habibovic2025} for recent work on classical considerations and the SFA). The real parts of the ionization and rescattering times of the first electron can be approximately related to the field extrema and zero crossings, respectively. Similarly, the ionization times of the second electron can be associated with the field extrema after rescattering has taken place.  Even though the real parts of the saddle-point solutions do not exactly correspond to the field maxima and zero crossings, the latter are useful for interpretational purposes, and allow one to easily classify different ionization and rescattering events. The insight into the dominance of specific events can be inferred from the imaginary part of the saddle-point solutions. All momentum distributions investigated in this paper are calculated using the saddle-point solutions, which are complex.
The saddle-point solutions are also used to compute partial momentum distributions for each electron. 

For the mapping of the ionization and rescattering times of the first electron, we employ the tangent construction, which is a graphical method relating the classical ionization and return times to field tangent near the field maxima and its intercept near a later field zero crossing, respectively\footnote{One should note that the classical ionization times given by the tangent construction is slightly after the maximum as times exactly at the maximum give a tangent of vanishing slope.}\cite{Faria1999}. A summary of the most important solutions, together with the fields used in this work, is provided in Fig.~\ref{fig:fields} and Table \ref{tab:mapping}. When calculating the partial electron momentum distributions, we neglect the prefactors and consider the $3s \rightarrow 3p$ transition for argon. The corresponding bound-state energies and ionization potentials are stated in Sec.~\ref{sec:model}.

\subsection{Field shapes and event mapping}
\label{sec: fieldshape}

In Fig.~\ref{fig:fields} we present the electric field (black solid line) and the corresponding vector potential (orange dashed line) as functions of time for the ($\omega$,$3\omega$) [panels (a), (b), and (c)], and ($\omega$,$2\omega$) [panels (d), (e), and (f)] bichromatic linearly polarized field. The ratio of the field amplitudes is $\xi=0.8$, and the relative phase is indicated in the panels. The ionization and rescattering times occur in pairs that coalesce at the boundaries of the classical allowed region \cite{Faria2003,Shaaran2010,Faria2012}. For that reason, we will use the notation $P_{n\mu}$, where $n$ is an integer and $\mu=a,b$, to refer to them depending on the half cycle taken into consideration.  The pairs of orbits for the first electron are indicated by arrows in Fig.~\ref{fig:fields}. Arrows in blue and cyan (red and pink) suggest that the contribution of a specific pair will populate the negative (positive) parallel momentum regions.  The second electron will be roughly freed at the subsequent electric field maxima, marked with shaded rectangles in the figure, whose colors were chosen to match those of the pairs associated with the first electron. Blue and cyan (red and pink) rectangles indicate the first and second ionization events of the second electron associated with the rescattering times in the negative (positive) momentum regions.  The most important orbits are denoted by $O_{n\mu}$, where $n$ is an integer and $\mu=a,b$ is associated with the half cycle from which it leaves. These orbits may lead to significant contributions for $p_{2||}>0$ or $p_{2||}<0$.
Gray dots indicate that the ionization event can be neglected because, for these solutions, the electric field is close to zero when the ionization happens. Throughout, we considered at most the two dominant pairs of solutions for the first electron, which are characterized by a large instantaneous amplitude $|E(t'')|$ and/or relatively short excursion times. The contributions of pairs with longer excursion times will be strongly suppressed due to wave-packet spreading orthogonal to the field, and, if a local maximum $|E(t'')|$ is much smaller than the absolute maximum of the field, ionization will decrease. For the second electron, we consider only ionization events occurring in the half cycle subsequent to rescattering, as later events will be rendered irrelevant due to bound-state depletion. This approximation has also been used in our previous publications \cite{Shaaran2010,Faria2012,Maxwell2015,Hashim2024}.  A summary of the relevant orbits is provided in Table \ref{tab:mapping}.

    \begin{table}[]
\begin{tabular}{c|c|c|c}
\begin{tabular}{ccc}\multicolumn{3}{c}{Fields}\\  $r$ &$s$ &$\phi$\end{tabular} & Events & Times (mod T) & $p_{\parallel}$ \\ \hline \hline 
\begin{tabular}{ccc} \\
   1  & 3& 0\\ \\
   1  & 3 & $\pi/2$\\ \\
   1&3&$-\pi/2$
\end{tabular}&  $e_1^-$ : \begin{tabular}{cc}
  $ P_{1a}$   & $P_{2a}$ \\
    $ P_{1b}$   & $P_{2b}$ 
\end{tabular}     
&   \begin{tabular}{c}
$-T/2 \leq t \leq 0$        \\
   $0\leq t \leq T/2$    
\end{tabular}          &     \begin{tabular}{c}
$>0$      \\
 $<0$
\end{tabular}         \\  
                                                             &     $e_2^-$ : \begin{tabular}{cc}
                                                             $ O_{1a}$    & $O_{2a}$  \\
                                                                 $ O_{1b}$ & $ O_{2b}$
                                                             \end{tabular}     &    \begin{tabular}{c}    $T/2\leq t \leq T$ 
                                                                   \\
                                                                  $T\leq t \leq 3T/2$ 
                                                             \end{tabular}        &         \begin{tabular}{c}
                                                                  $>0$, $<0$\\
                                                                  $<0$, $>0$
                                                             \end{tabular}       
                \\ \hline
\begin{tabular}{ccc} \\
   1  & 2& 0\\ \\
   1  & 2& $3\pi/4$\\ 
\end{tabular}&      $e_1^-$ : \begin{tabular}{cc}
  $ P_{1a}$  \\
    $ P_{1b}$   & $P_{2b}$ 
\end{tabular}        &          \begin{tabular}{c}
$-T/2 \leq t \leq 0$        \\
   $0\leq t \leq T/2$    
\end{tabular}        &      \begin{tabular}{c}
$>0$      \\
 $<0$
\end{tabular}     \\ 
                                                             &     $e_2^-$ : \begin{tabular}{cc}
                                                             $ O_{1a}$    \\
                                                                 $ O_{1b}$ & $ O_{2b}$
                                                             \end{tabular}      &    
                                                             \begin{tabular}{c}    $T/2\leq t \leq T$ 
                                                                   \\
                                                                  $T\leq t \leq 3T/2$ 
                                                             \end{tabular} &  \begin{tabular}{c}
                                                                  $0$\\
                                                                  $<0$, $>0$
                                                             \end{tabular}  
                                                              \\ \hline

\multirow{2}{*}{\hspace{-0.5cm} 1 2 \hspace*{0.2cm}$\pi$}                                            &    $e_1^-$ : \begin{tabular}{cc}
  $ P_{1a}$  \\
    $ P_{1b}$  
\end{tabular}        &          \begin{tabular}{c}
$-T/2 \leq t \leq 0$        \\
   $0\leq t \leq T/2$    
\end{tabular}        &      \begin{tabular}{c}
$>0$      \\
 $<0$
\end{tabular}                \\  
                                                             &     $e_2^-$ : \begin{tabular}{cc}
                                                             $ O_{1a}$    &   \\
                                                                 $ O_{1b}$&
                                                             \end{tabular}    &     \begin{tabular}{c}    $T/2\leq t \leq T$ 
                                                                   \\
                                                                  $T\leq t \leq 3T/2$ 
                                                             \end{tabular}           & \begin{tabular}{c}
                                                                  $>0$\\
                                                                  $>0$
                                                             \end{tabular}  \\ \hline \hline
\end{tabular}
    \caption{Relevant events for the first and second electron, for the driving fields employed in Fig.~\ref{fig:fields}. The first column gives the field parameters, the second column gives the events (pairs and orbits) associated with the first and second electron, respectively, the third column gives the time interval for which these events occur, and the last column states the signs of the parallel momenta associated with each event. }
    \label{tab:mapping}
\end{table}
 Figure~\ref{fig:fields} shows that, for a two-color field, there are key differences from the behavior observed for a monochromatic field or few-cycle pulses. First, within a field half-cycle, there may be more than one ionization event leading to rescattering at the same field zero crossing. For instance, in the upper row of Fig.~\ref{fig:fields}, there are two pairs $P_{1a,b}$ and $P_{2a,b}$, with the ionization and return times occurring at different and subsequent half cycles. The pairs $P_{1a}$ and $P_{2a}$ ($P_{1b}$ and $P_{2b}$) populate the positive (negative) momentum regions. Besides the sign reversal in the momentum, the dynamics unleashed by these pairs are identical. This is expected due to the half-cycle symmetry that exists for the $(\omega,3\omega)$ field.  In the lower row, we see a single event associated with the pair  $P_{1a}$, and the two pairs $P_{1b}$ and $P_{2b}$ occur only every second half-cycle as the $(\omega,2\omega)$ field is not half-cycle symmetric. Similarly, in the half cycle subsequent to the rescattering, there are up to two field maxima for which the second electron may be freed.  Also, for pairs of events located in the same half cycle, the field extrema at the time of ionization do not correspond to zero crossings of the vector potential, which will have implications for the most probable momentum with which the second electron will reach the continuum.  In contrast, for a monochromatic field \cite{Maxwell2015,Maxwell2016} or few-cycle pulses \cite{Shaaran2012,Faria2012,Hashim2024}, there is only one pair of events per half cycle and the field extrema exactly or approximately correspond to zero crossings of $A(t)$, respectively. 

Changing the relative phase between the two driving waves will influence the most relevant pair, as exemplified in Fig.~\ref{fig:fields}. Figure~\ref{fig:fields}(a) shows that, for $(\omega,3\omega)$ field with the relative phase $\phi=0$, the two field extrema associated with the pairs $P_{1a,b}$ and $P_{2a,b}$ have equal magnitude. Furthermore, both pairs have ionization times close to each other and the same rescattering time. This implies that the electron excursion times in the continuum will be similar. Therefore, we expect their contributions to be comparable.  The same holds for the ionization events associated with the second electron: they are expected to yield comparable contributions, which, however, will be located in opposite momentum regions. This can be inferred by the instantaneous value of $-A(t)$, where, according to the saddle-point equation \eqref{eq:spati}, the parallel momentum distribution is centered. For the events marked with the blue and pink (red and cyan) shaded rectangles, $A(t)$ is positive (negative), which means that the most probable momentum $p_{2\parallel}$ associated with this event will be negative (positive). Because both field extrema are equal in magnitude, we can infer that the positive and negative momentum regions for $p_{2\parallel}$ will be equally occupied. This field shape corresponds to a scenario described in \cite{Rook2022}, for which the three symmetries that exist for a monochromatic field are also present for a $(\omega,3 \omega)$ field. 

In Figs.~\ref{fig:fields}(b) and (c), the two field extrema have been made unequal by changing $\phi$. Hence, although the half-cycle symmetry is retained, the reflection symmetries about the field extrema and zero crossings are broken.  For $\phi=\pi/2$ [Fig.~\ref{fig:fields}(b)], the field extrema associated with $P_{1a,b}$ have increased in magnitude, with regard to those associated with $P_{2a,b}$.  Therefore, we expect the pairs $P_1$ to be dominant. A similar argument can be applied to the second electron: the events immediately after rescattering have lost their significance, while the later events have become more important.  This is due to the decrease and increase in the corresponding field extrema, respectively.  One should note that, for consecutive half cycles, opposite momentum regions will be populated. For instance, the ionization events for $T/2 \leq t \leq T$ ($T \leq t \leq 3T/2$) will lead to predominantly $p_{2\parallel}<0$ ($p_{2\parallel}>0$), because the vector potential is positive (negative) at the time of the dominant event. For $\phi=-\pi/2$ [Fig.~\ref{fig:fields}(c)], the scenario has been reversed and the field extremes associated with $P_2$ were made more prominent. This renders this pair dominant for the first electron, although it corresponds to a slightly longer orbit. Likewise, the first field extremes after the crossing are now associated with the dominant events for the second electron. For subsequent half cycles, the signs of the region occupied by $p_{2\parallel}$ will alternate, but they will be the opposite of what was observed for $\phi=\pi/2$. Explicitly, if the ionization events happened at  $T/2 \leq t \leq T$ ($T \leq t \leq 3T/2$), then  $p_{2\parallel}>0$ ($p_{2\parallel}<0$). Once more, this can be inferred from the sign of the vector potential at the time of the dominant event. 

The situation becomes slightly more complicated for the $(\omega$,$2\omega$) field due to the absence of the half-cycle symmetry. Still, the conclusions drawn from the previous case are largely applicable, but with the periodicity of a full cycle. For $\phi=\pi/2$ [Fig.~\ref{fig:fields}(d)], we see that, for  $0 \leq t \leq T/2$ modulo $T$ the field has two maxima symmetric around the minimum at around $T/4$ (mod $T$), while for $T/2 \leq t \leq T$ modulo $T$ there is a single extremum. This means that there is the reflection symmetry  $\mathcal{T}_R\left(\tau_{ex}\right)$ about the field extremes, but the symmetry $\mathcal{F}\mathcal{T}_R\left(\tau_{cr}\right)$ about the crossings is broken, in addition to the half-cycle symmetry.  The dominant events for the first electron are related to the pair $P_{1a}$, highlighted by the red arrow starting around $t=-T/4$. In the subsequent half cycle, there are two symmetric field maxima that resemble those identified for the $(\omega,3\omega)$ field but are temporally further apart. They are associated with the pairs  $P_{1b}$ and $P_{2b}$, for which the electron returns near $T$. Because of the shorter electron excursion amplitude, the contributions from $P_{1b}$ prevail. The scattering event associated with $P_{1a}$ is followed by the ionization event near  $t=3T/4$ for the second electron. For this event, $A(t) \simeq 0$, which means that the most probable momentum for the second electron will be vanishing. The other two ionization events, after the rescattering triggered by $P_{1}$, are similar to those assessed for the $(\omega, 3\omega)$  field, and will lead to probability densities centered at nonvanishing momenta $p_{2\parallel}$ of opposite signs. 

Changing the relative phase $\phi$ disrupts this pattern by breaking the reflection symmetry around the field extremes. For instance, in Figs.~\ref{fig:fields}(e) and (f) [$\phi=3\pi/4$ and $\phi=\pi$, respectively], we  suppress the field peaks associated with $P_{2b}$, until, for $\phi=\pi$, they are vanishingly small. Furthermore, the ionization events after the rescattering around $T$ will now have different prominence, with that close to the zero crossing losing relevance, until, for $\phi=\pi$, its contributions become negligible [Fig.~\ref{fig:fields}(f)]. Another, subtler effect is that, for the other half cycles, although $A(t)=0$ for the ionization events happening after the recollision caused by $P_{1a}$, the gradients are no longer symmetric. This will affect the resulting electron momentum distributions. Finally, it is noteworthy that, for $\phi=\pi$, the symmetry $\mathcal{F}\mathcal{T}_R\left(\tau_{cr}\right)E(t)=E(t)$ around the field zero crossings holds. Furthermore, the vector potential is reflection symmetric around its maxima.  

\begin{figure*}[!htbp]
\centering
\includegraphics[width=0.97\textwidth]{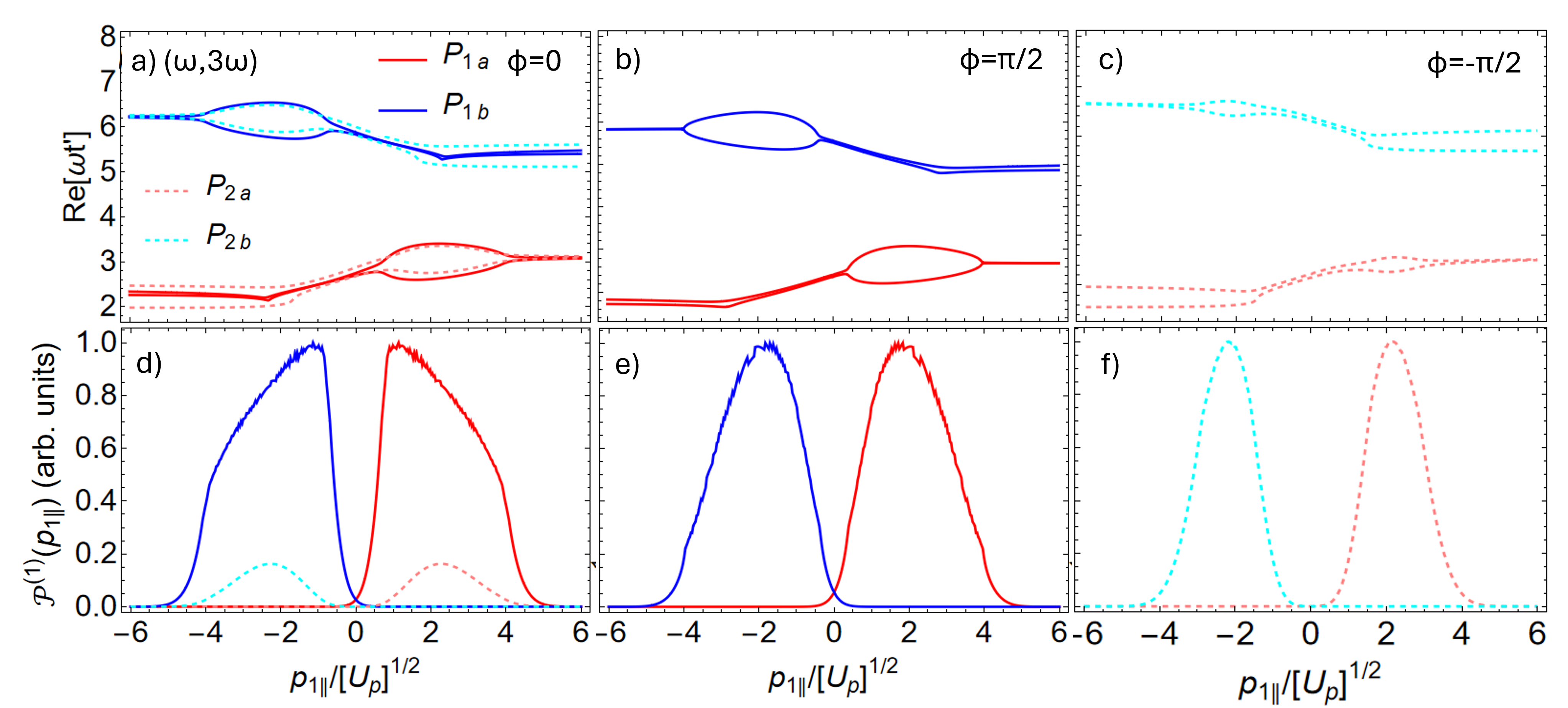}
    \caption{Real parts of the rescattering time calculated for $p_{1\perp}=0$ using the saddle-point equations [upper row; panels (a), (b), and (c)] and the partial momentum distribution of the first electron given by Eq.~\eqref{eq:partialdistr} [lower row; panels (d), (e), and (f)] as functions of the parallel momentum $p_{1||}$, for
    values of the driving-field parameters as in the upper row of Fig.~\ref{fig:fields} $(\omega,3\omega)$. The colors of the saddle-point solutions and partial distributions have been chosen to match those in Fig.~\ref{fig:fields}. Specifically, the contributions associated with $P_{1a}$ and $P_{1b}$ are plotted using solid red and blue lines, respectively, while those related to $P_{2a}$ and $P_{2b}$ are displayed using dashed pink and cyan lines. 
    Only the pairs with dominant contributions to the photoelectron yield are taken into consideration, except for the field with $\phi=0$ in which case,  one additional pair is considered. The partial momentum distribution for this pair has been scaled by 10. The values of the relative phase are indicated in the legends. The intensity of the $\omega$ field component is $E_{\omega}^2=6\times 10^{13}$W/cm$^2$ and the fundamental wavelength is 800~nm.}
    \label{fig:1ew3wpart}
\end{figure*}

\begin{figure*}[!htbp]
\centering
\includegraphics[width=0.97\textwidth]{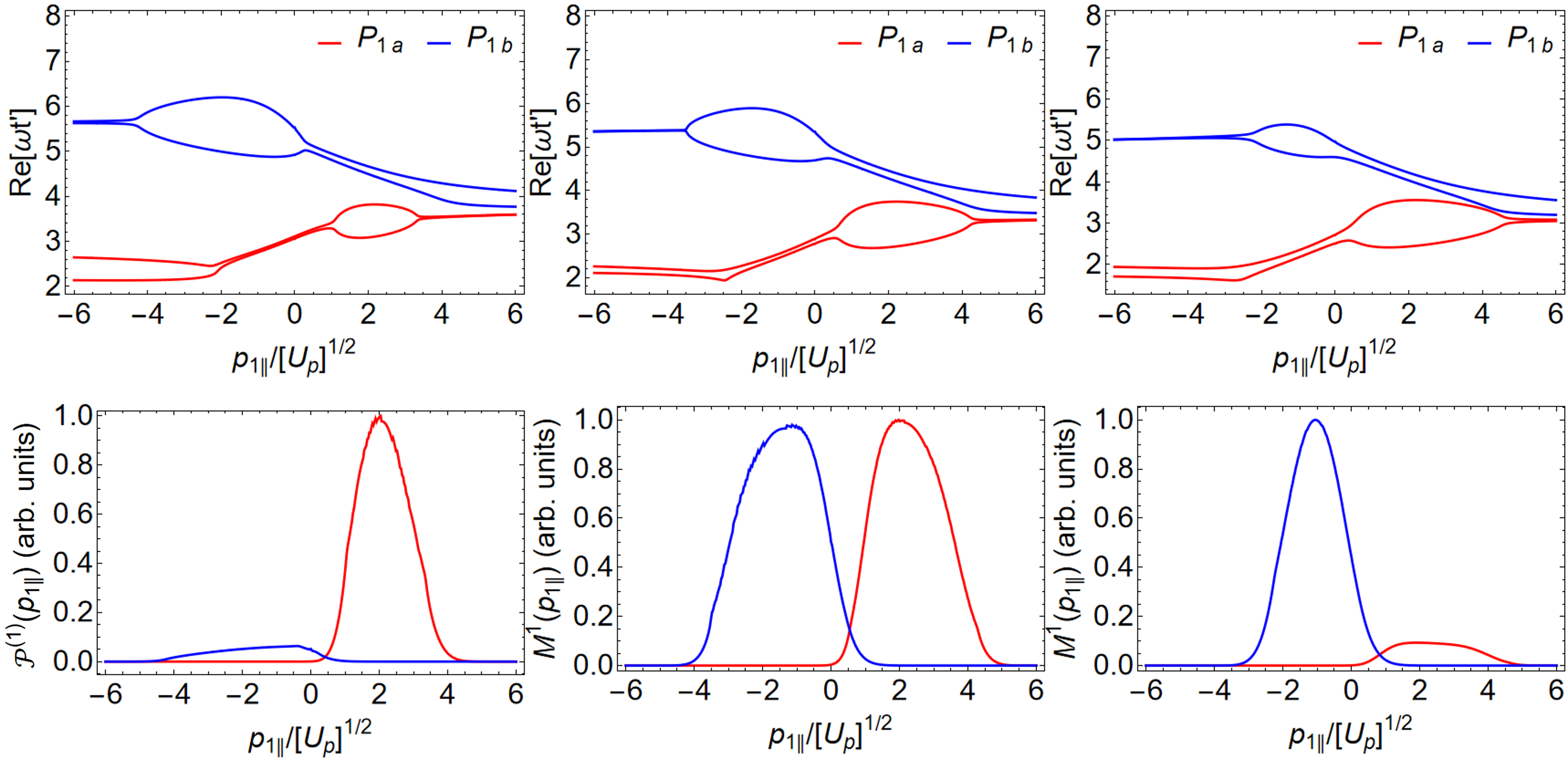}
    \caption{Real parts of the rescattering time calculated for $p_{1\perp}=0$ using the saddle-point equations [upper row; panels (a), (b), and (c)] and the partial momentum distribution of the first electron given by Eq.~\eqref{eq:partialdistr} [lower row; panels (d), (e), and (f)] as functions of the parallel momentum $p_{1||}$, for the values of the driving-field parameters as in the lower row of Fig.~\ref{fig:fields} $(\omega,2\omega)$. Only the pairs with dominant contributions to the photoelectron yield, namely $P_{1a}$ and $P_{1b}$ are taken into consideration, and their corresponding saddle-point solutions and probability densities are plotted using solid red and blue lines, respectively. The values of the relative phase are indicated in the legends. The intensity of the $\omega$ field component is $E_{\omega}^2=6\times 10^{13}$W/cm$^2$ and the fundamental wavelength is 800~nm.}
    \label{fig:1ew2wpart}
\end{figure*}
\subsection{Partial momentum distributions - first electron}
\label{sec:partial1st}
In Fig.~\ref{fig:1ew3wpart}, we analyze the contributions of the saddle-point solutions to the first-electron partial RESI transition probability for the same $(\omega$,$3\omega$) driving fields as in Fig.~\ref{fig:fields}. In the upper row  [panels (a), (b), and (c)], we present the real part of the rescattering time as a function of the parallel momentum $p_{1||}$ computed using the corresponding saddle-point equations for orthogonal momentum  $p_{1\perp}=0$. 
In the lower row  [panels (d), (e), and (f)], we plot the corresponding partial momentum distributions. They have been calculated without any prefactors to avoid additional momentum biases. Here, we consider only the dominant pairs, and follow the notation used in Fig.~\ref{fig:fields} and Table \ref{tab:mapping}. The pairs $P_{2a,b}$ ($P_{1a,b}$) and their contributions are not plotted for $\phi=\pi/2$ ($\phi=-\pi/2$), as their contributions are orders of magnitude smaller than those of the other events. For that reason, we do not include these in any of the momentum-distribution calculations.

The sets of times plotted in Figs.~\ref{fig:1ew3wpart}(a), (b), and (c) occur in pairs that nearly coalesce at a minimum and a maximum value of $p_{1\parallel}$. Those momenta mark the boundary of the region for which rescattering has a classical counterpart, which we refer to as the classically allowed region. They are also roughly centered at $p_{1 \parallel}=-A(t')$, which gives the most probable momentum associated with rescattering. The solutions associated with the pairs $P_{1a}$, $P_{2a}$ and $P_{1b}$, $P_{2b}$ are displaced by half a cycle and are the mirror image of each other regarding $p_{1\parallel}=0$.  This behavior follows from the $(\omega,3\omega)$ field being half-cycle symmetric.
For relative phase $\phi=0$, the classically allowed region is similar for the pairs $P_1$ and $P_2$ due to the other two symmetries associated with the field extremes and zero crossings, namely $\mathcal{T}_R\left(\tau_{ex}\right)E(t)=E(t)$ and $\mathcal{F}\mathcal{T}_R\left(\tau_{cr}\right)E(t)=E(t)$, being present. These symmetries are broken for the other two relative phases. Fig.~\ref{fig:1ew3wpart}(a) also shows that the rescattering times associated with $P_{2b}$ (blue solid line) and with  $P_{1b}$ (dashed cyan line) are very close. This is consistent with the classical mapping performed in the previous section, which indicates that, for the two events, the electron should return near the same field zero crossing. The same holds for the pairs $P_{2a}$ (dashed pink line) and $P_{1a}$ (red solid line), displaced by half a cycle. For large enough absolute values of $p_{1\parallel}$, the solutions $\mathrm{Re}[\omega t']$ tend to a value close to the zero crossing. This behavior has also been observed for the field parameters in \ref{fig:1ew3wpart}(b), and (c), although, in these latter cases, only the dominant saddle-point solutions have been included. The remaining solutions lead to contributions that are several orders of magnitude smaller, and thus are not relevant for th present discussion. 

The partial electron momentum distributions, displayed in Figs.~\ref{fig:1ew3wpart}(d), (e), and (f), are invariant regarding the transformation $p_{1||}\rightarrow -p_{1||}$ (cf. the blue and red solid and the cyan and pink dashed lines in the lower row of Fig.~\ref{fig:1ew3wpart}), being the mirror image of the other. This is expected, due to the half-cycle symmetry, and is also a consequence of the imaginary parts of the ionization times $t''$, which are related to the probability that each electron tunnels through the instantaneous potential barrier that is narrowest near the field extrema\footnote{The probability of an electron tunneling through a potential barrier and reaching a continuum state is proportional to $\exp[-2\mathrm{Im}[S]$, where $S$ is the semiclassical action. Because its dominant term is proportional to $U_p t$, one can, to first approximation, assert the dominance of an event, linked to a specific quantum pathway, by inspecting $\mathrm{Im}[t]$ (for a review see \cite{Popruzhenko2014a}) }, also being symmetric upon  $p_{1||}\rightarrow -p_{1||}$.
 Figures~\ref{fig:1ew3wpart}(d), (e), and (f) also provide insight into the dominance of a specific pair, determined by the interplay between the instantaneous tunneling probability of the first electron, its excursion time in the continuum and the classically allowed region determined at rescattering. For instance,  for Fig.~\ref{fig:1ew3wpart}(d), the tunneling probabilities associated with the pairs $P_{1a}$ and $P_{2a}$ are equal, and the corresponding classically allowed regions are similar. The same holds for those of pairs $P_{1b}$ and $P_{2b}$, which are displaced by half a cycle. 
Nonetheless, the partial momentum distribution which corresponds to the pairs $P_{2a,b}$ is scaled by 10, which means that the contribution of this pair is approximately one order of magnitude smaller than the contribution of the pairs $P_{1a,b}$. This is due to the excursion amplitude being larger for $P_2$. If the tunneling  probabilities associated with $P_{1a,b}$ are made larger, for example, by taking $\phi=\pi/2$ [Fig.~\ref{fig:1ew3wpart}(e)], the contributions of $P_{2a,b}$ are rendered vanishingly small. Figure~\ref{fig:1ew3wpart}(f) shows how the tunneling probability trumps the excursion amplitude in determining the relevance of an event. By increasing the field maxima associated with $P_{2a,b}$, these pairs of orbits were made dominant despite the electron spending a longer time in the continuum.  

A similar study can be performed for the ($\omega$,$2\omega$) driving field, considering the dominant pairs as stated in Table \ref{tab:mapping}. In Fig.~\ref{fig:1ew2wpart} we present the real part of the rescattering time as a function of the parallel momentum $p_{1||}$ (upper row) for pairs of the saddle-point solutions associated with $P_{1\mu}$, $\mu=a,b$, in the lower row of Fig.~\ref{fig:fields}, calculated for orthogonal momentum  $p_{1\perp}=0$. In the lower row of Fig.~\ref{fig:1ew2wpart}, we plot the corresponding partial momentum distribution of the first electron [given by Eq.~\eqref{eq:partialdistr}] calculated as a function of the parallel momentum $p_{1||}$ without any prefactors. The $(\omega$,$2\omega)$ field does not possess the half-cycle symmetry so that the saddle-point solutions in the $p_{1||}>0$ and  $p_{1||}<0$ parts of the momentum plane are not related via a simple translation or reflection. Consequently, the partial momentum distributions in these momentum regions are different. 
Furthermore, for the $(\omega,2\omega)$ field with the relative phase $\phi=0^\circ$, the contributions of pair $P_{1a}$ dominate over those of $P_{1b}$ [see Fig.~\ref{fig:1ew2wpart}(d)], although this pair is associated with a longer excursion time and a much smaller classically allowed region [see Fig.~\ref{fig:fields}(d) and Fig.~\ref{fig:1ew2wpart}(a), respectively]. This is evidence that a higher tunneling probability supersedes those two other criteria. The picture is approximately swapped for the results displayed in Fig.~\ref{fig:1ew2wpart}(f) [$(\omega,2\omega)$ field with the relative phase $\phi=\pi$], for which the contributions of $P_{1b}$ prevail. An inspection of Fig.~\ref{fig:fields}(f) shows that the tunneling probabilities, associated with the field extrema, are comparable for both $P_{1a}$ and $P_{1b}$, but the electron excursion times associated with $P_{1b}$ are shorter. This outweighs the larger classically allowed region observed for $P_{1a}$ [see Fig.~\ref{fig:1ew2wpart}(c)]. 
For $\phi=3\pi/4$ [Fig.~\ref{fig:1ew2wpart}(e)] the first-electron partial momentum distributions are approximately the same in both parts of the momentum plane. This is consistent with the real parts of the rescattering times, which are roughly mirror symmetric about $p_{1\parallel}=0$. This accidental symmetry is likely due to the classically allowed regions being very similar, and the other, more important contributing factors roughly compensating each other. According to Fig.~\ref{fig:fields}(e), the instantaneous tunneling probability associated with $P_{1a}$ is larger, due to a larger absolute value of $E(t'')$, but the excursion time for $P_{1b}$ is much shorter. 
 
\begin{figure*}[!htbp]
\centering
\includegraphics[width=0.97\textwidth]{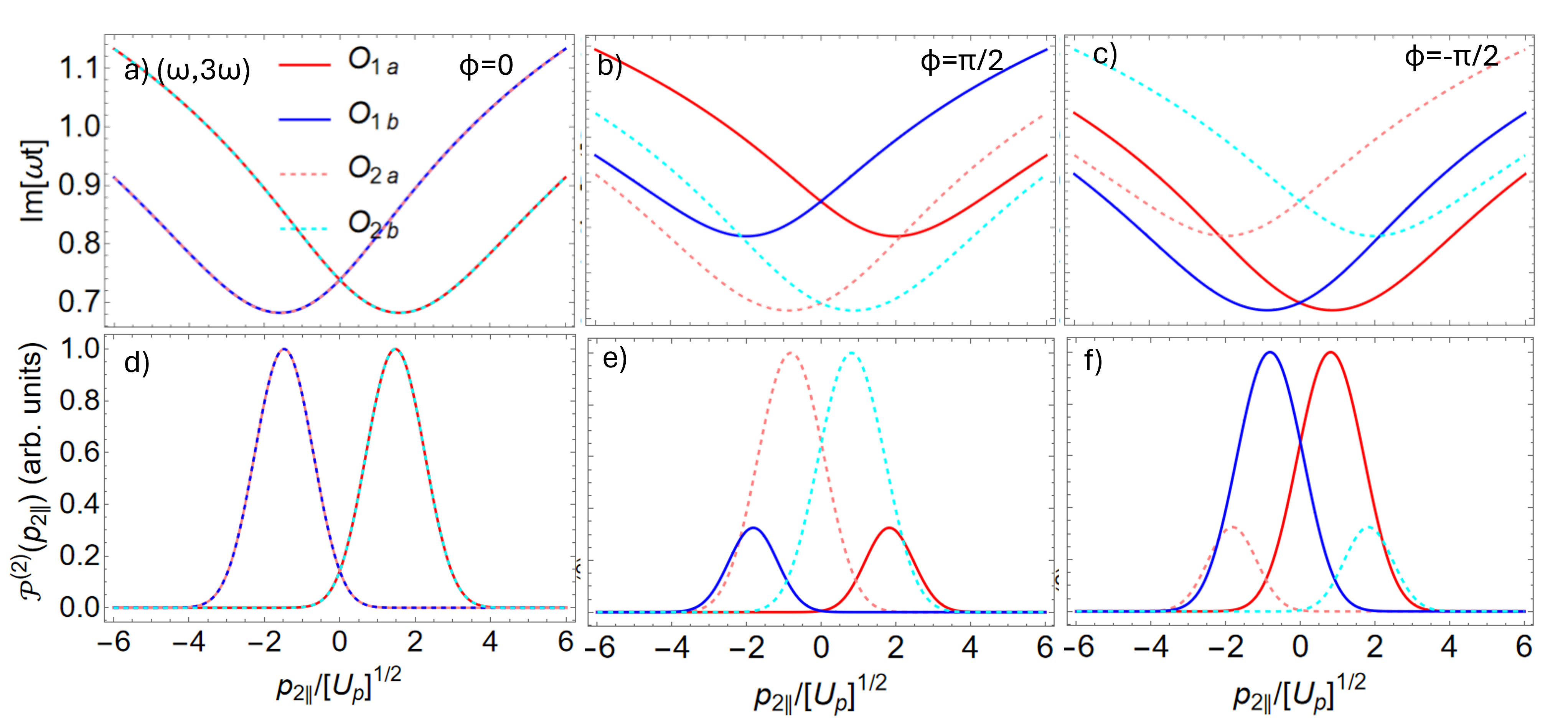}
    \caption{Imaginary part of the ionization time $t$ calculated for $p_{2\perp}=0$ [upper row; panels (a), (b), and (c)] and the partial momentum distribution of the second electron given by Eq.~\eqref{eq:partialdistr} [lower row; panels (d), (e), and (f)] as functions of the parallel momentum $p_{2||}$, for the values of the driving-field parameters as in the upper row of Fig.~\ref{fig:fields} ($\omega$,$3\omega$).
     The field intensity and the wavelength of the fundamental are the same as in Fig.~\ref{fig:1ew3wpart}.
    Only the solutions with non-negligible contributions are taken into consideration. The colors of the lines correspond to those of the shaded rectangles in the upper panels of Fig.~\ref{fig:fields}. To distinguish between different events, the contributions associated with $O_{1a}$ and $O_{1b}$ are plotted using solid red and blue lines, respectively, while those related to $O_{2a}$ and $O_{2b}$ are displayed using pink and cyan dashed lines. }
    \label{fig:2ew3wpart}
\end{figure*}
\subsection{Partial momentum distributions - second electron}
\label{sec:partialsecond}

We now turn our attention to the second electron, which may tunnel from an excited state some time after the recollision of the first electron.  Within the saddle-point framework, its ionization amplitude is related to the imaginary part of the ionization time so that an increase in the value of the imaginary part of the ionization time $t$ leads to an exponential decrease in the ionization amplitude. Therefore, we plot $\mathrm{Im}[t]$ in the figures that follow, together with the corresponding partial probability distributions. 

In Fig.~\ref{fig:2ew3wpart} we present the imaginary part of the ionization time of the second electron calculated for $p_{2\perp}=0$ [upper row; panels (a), (b), and (c)] and the partial momentum distribution of the second electron [lower row; panels (d), (e), and (f)] as functions of the parallel momentum $p_{2||}$, for the values of the driving-field parameters as in the upper row of Fig.~\ref{fig:fields} [i.e., for the ($\omega$,$3\omega$) field], and for the saddle-point solutions associated with the events $O_{na,b}$, $n=1,2$ in Table \ref{tab:mapping}. The times associated with each event are $t_{na,b}$, $n=1,2$, following the same notation, i.e., $t_{1a}$ corresponds to $O_{1a}$, and so forth. 
A common feature observed in Figs.~\ref{fig:2ew3wpart}(a), (b), and (c) is that the minima of $\mathrm{Im}[t]$ occur at nonvanishing parallel momenta, which are symmetric about $p_{2\parallel}=0$. These momenta are approximately given by $p_{2\parallel}=\mp A(t)$, and agree with the location of the maxima of the partial electron momentum distributions [see Figs.~\ref{fig:2ew3wpart}(d), (e), and (f)]. Furthermore, the saddle-point solutions and partial momentum distributions displaced by half a cycle are mirror symmetric. This stems from the half-cycle symmetry of the ($\omega$,$3\omega$) field. Specifically, the contributions of $O_{1a}$ and $O_{2b}$ are peaked in the positive parallel momentum region, while those of $O_{1b}$ and $O_{2a}$ have maxima for negative $p_{2\parallel}$.

Nonetheless, the dominant orbits and the momentum regions they occupy depend on the relative phases $\phi$.
For $\phi=0$ [Fig.~\ref{fig:2ew3wpart}(a)], $\mathrm{Im}[t_{1a}]=\mathrm{Im}[t_{2b}]$ and $\mathrm{Im}[t_{1b}]=\mathrm{Im}[t_{2a}]$, with $\mathrm{Im}[t_{1a,b}]$ being the mirror image of $\mathrm{Im}[t_{2a,b}]$. The same symmetries hold for the partial electron momentum distributions, shown in Fig.~\ref{fig:2ew3wpart}(d), with the contributions of $O_{1a}$ and $O_{2b}$ ($O_{1b}$ and $O_{2a}$) being identical.  Physically, these additional features are due to the symmetries $\mathcal{T}_R\left(\tau_{ex}\right)E(t)=E(t)$ and $\mathcal{F}\mathcal{T}_R\left(\tau_{cr}\right)E(t)=E(t)$ around the field extremes and zero crossing being present in addition to the half-cycle symmetry. 

In contrast, for the relative phase $\phi=\pi/2$, these additional symmetries are broken, and dominant contributions to the partial momentum distribution of the second electron come from the solutions $O_{2a}$ and $O_{2b}$. These solutions have the smallest imaginary part of the ionization time [see the pink and cyan dashed lines in Figs.~\ref{fig:2ew3wpart}(b) and (e)]. The partial momentum distributions which correspond to the solutions $O_{1a}$ and $O_{1b}$  are scaled by 10 in Fig.~\ref{fig:2ew3wpart}(e) which means that only two saddle-point solutions have to be taken into consideration. This is expected since the electric-field peaks related to the solutions $O_{2a}$ and $O_{2b}$ are much stronger than the peaks related to the solutions $O_{1a}$ and $O_{1b}$  [cf. the shaded rectangles in Fig.~\ref{fig:fields}(b) and their corresponding field amplitudes]. Finally, for the relative phase $\phi=-\pi/2$ the situation is reversed with respect to the times and the distributions displayed in Figs.~\ref{fig:2ew3wpart}(b)  and (e). The dominant contributions to the second-electron yield come from the saddle-point solutions $O_{1a}$ and $O_{1b}$  [red and blue solid lines in Figs.~\ref{fig:2ew3wpart}(c)  and (f)], while the contributions of $O_{2a}$ and $O_{2b}$  [pink and cyan dashed lines in Figs.~\ref{fig:2ew3wpart}(c)  and (f)] are much weaker. In Fig.~\ref{fig:2ew3wpart}(f),  the contributions of $O_{2a}$ and $O_{2b}$  are  scaled  by the factor 10. 

\begin{figure*}[!htbp]
\centering
\includegraphics[width=0.97\textwidth]{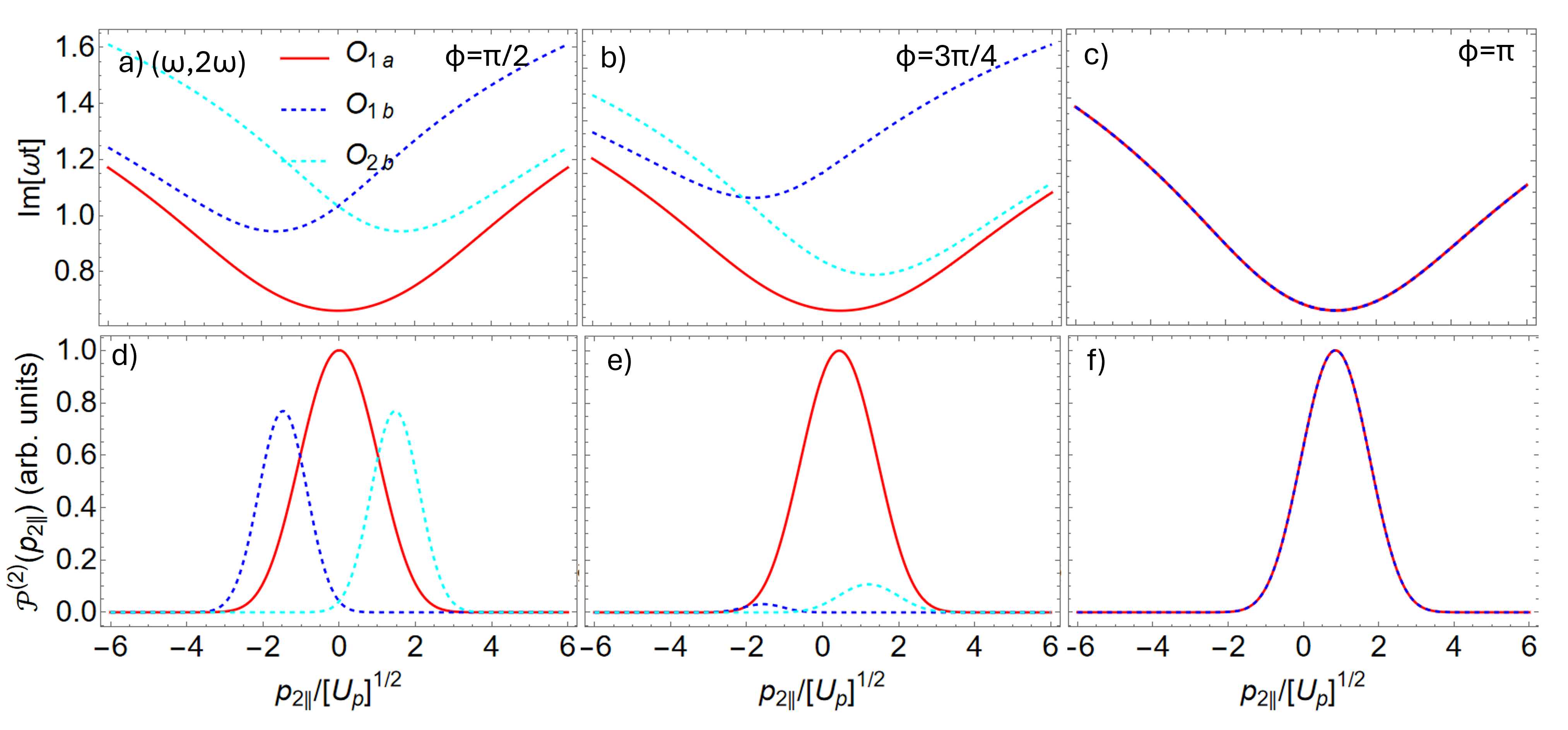}
    \caption{Imaginary part of the ionization time $t$ calculated for $p_{2\perp}=0$ [upper row; panels (a), (b), and (c)] and the partial momentum distribution of the second electron given by Eq.~\eqref{eq:partialdistr} [lower row; panels (d), (e), and (f)] as functions of the parallel momentum $p_{2||}$, for the values of the driving-field parameters as in the lower row of Fig.~\ref{fig:fields} ($\omega$,$2\omega$). The field intensity and the wavelength of the fundamental are the same as in Fig.~\ref{fig:1ew3wpart}. Only the solutions with nonnegligible contributions are taken into consideration. The colors of the lines correspond to those of the shaded rectangles in the lower panels of Fig.~\ref{fig:fields}. However, the styles and the shades are slightly changed to facilitate the discussion, with the contributions of $O_{1a}$  plotted with red solid lines, those of  $O_{1b}$ with blue dashed lines, and those of $O_{2b}$ with cyan dashed lines. Note that this convention is different from that used in the previous figure. }\label{fig:2ew2wpart}
\end{figure*}
A similar analysis can be performed for the $(\omega$,$2\omega$) field. In this case, there are four saddle-point solutions per cycle, but usually three and sometimes only two lead to a significant contribution to the second-electron yield. Furthermore, the half-cycle symmetry is broken, which means that the solutions obtained for the first half-cycle are not the mirror image of those of the second. Here, we consider the solutions outlined in Table \ref{tab:mapping}. In all cases, there is a single event $O_{1a}$ for times $(2n-1)T/2 \leq t \leq nT$, and at most two relevant events for the subsequent half cycles, i.e., $nT \leq t \leq (2n+1)T/2$, here called $O_{1b}$ and $O_{2b}$. The results analogous to those presented in Fig.\ref{fig:2ew3wpart} but for the ($\omega$,$2\omega$) field are shown in Fig.~\ref{fig:2ew2wpart}. 

For $\phi=\pi/2$ and $\phi=3/4 \pi$ [first and second columns of Fig.~\ref{fig:2ew2wpart}], $O_{1a}$ is the dominant solution [see the red solid lines in Fig.~\ref{fig:2ew2wpart}(d) and (e)], with the smallest imaginary part for $t_{1a}$ and the largest partial electron momentum distribution. The other two solutions lead to much less relevant contributions  [the blue and cyan dashed lines in Fig.~\ref{fig:2ew2wpart}(d) are scaled by 100 and their counterparts in Fig.~\ref{fig:2ew2wpart}(e) are even smaller]. This is in accordance with the electric-field profiles shown in Figs~\ref{fig:fields}(d) and (e). In particular, the peak that corresponds to the solution $O_{1a}$ [denoted by the red rectangles in Figs~\ref{fig:fields}(d) and (e)] is the strongest thus leading to the most prominent partial contribution, while the solutions $O_{1b}$ and $O_{2b}$ correspond to the much smaller peaks. The minimum of  $\mathrm{Im}[t_{1a}]$ and the peaks of the partial momentum distributions are located around  $p_{2\parallel}=0$. This is expected as, for $O_{1a}$, $A(t_{1a})\simeq 0$.  The remaining solutions exhibit minima at non-vanishing momenta and behave like their counterparts in the $(\omega,3\omega)$ case.  For $\phi=\pi/2$ [first column in Fig.~\ref{fig:2ew2wpart}], the imaginary parts of the ionization times related to $O_{1b}$ and $O_{2b}$ and the corresponding partial electron momentum distributions are the mirror image of each other, while, for $O_{1a}$, the partial distribution and $\mathrm{Im}[t_{1a}]$ is perfectly symmetric around $p_{2\parallel}=0$. These behaviors are explained by the field being reflection-symmetric around its maxima. For $\phi=3/4\pi$, this no longer holds and the gradients around the maxima are unequal. Therefore, the contributions of $O_{1a}$ are slightly skewed towards $p_{2||}>0$ and those of $O_{1b}$ and $O_{2b}$  are no longer mirror symmetric about $p_{2\parallel}=0$.  Finally, for  $\phi=\pi$, the two solutions $O_{1a}$ and $O_{2b}$ have the same imaginary part of the ionization time, thus leading to identical partial momentum distributions. This is due to the symmetry $\mathcal{F}\mathcal{T}_R\left(\tau_{cr}\right)E(t)=E(t)$ around the crossings being present. Furthermore, the vector potential is reflection symmetric about its maxima and minima, which guarantees the same momentum transfer for both events.  

\section{Two-electron momentum distributions}
\label{section:momdists}

We now investigate the RESI two-electron momentum distributions in the $p_{1||}p_{2||}$ plane. First, using the knowledge obtained in the previous section, we sketch the shapes of these distributions for the driving fields considered in this work. These diagrammatic representations are presented in Fig.~\ref{fig:diagrams}, and consider the dominant events determined in the previous section. They are helpful to assess the shapes of the fully incoherent distributions and also to map the key interference events. They constitute a modified version of those in our previous work \cite{Hashim2024}, as, depending on the circumstances, one needs to account for more than one event per half cycle. A straight line corresponds to the contribution of a specific event to the distribution. The length and width of each horizontal line are determined by the kinematic constraints associated with the first and second electron, respectively, but these roles are reversed once the electron momenta are exchanged. Different and the same colors are associated with different and the same event, respectively. We recall that the events corresponding to the first electron are denoted by $P_{ij}$, while the orbits for the second electron are denoted by $O_{kl}$ where $i,k$ are integers and $j,l$ are either $a$ or $b$. Due to the depletion of the bound state, the prevailing ionization times of the second electron occur within the half-cycle after the rescattering of the first electron. This means that, for the $(\omega$, $3\omega)$ field, the contributions of the $P_{1a}$ pair for $p_{1||}>0$ should be combined with the $O_{1a}$ and $O_{2a}$ solutions for the second electron, while the contributions of the $P_{1b}$ pair for $p_{1||}<0$ should be combined with the $O_{1b}$ and $O_{2b}$ solutions for the second electron. For the $(\omega$, $2\omega)$ driving field, the $P_{1a}$ pair should be combined with $O_{1a}$, and the $P_{1b}$ pair with $O_{1b}$ and $O_{2b}$.

The diagram in Fig.~\ref{fig:diagrams}(a) is associated with the $(\omega,3\omega, \phi=0)$ field. The RESI distributions are expected to be fourfold symmetric, with a single event contributing to the probability density in the first and third quadrants ($P_{1a}O_{1a}$ and $P_{1b}O_{1b}$, respectively) of the $p_{1\parallel}p_{2\parallel}$ plane, and two events ($P_{1a}O_{2a}$ and $P_{1b}O_{2b}$) determining the probability distribution in the second and fourth quadrants. An interesting feature is that, because the most probable parallel momenta are nonvanishing for the second electron, the distributions are not expected to occupy the $p_{n\parallel}=0$ axes. For $(\omega,3\omega, \phi=\pi/2)$  and $(\omega,3\omega, \phi=-\pi/2)$ fields [Figs.~\ref{fig:diagrams}(b) and (c), respectively], we anticipate the contributions in the first and third (second and fourth) quadrants to be suppressed. A noteworthy feature is that, upon symmetrization, contributions occupying the second quadrant will move to the fourth and vice versa, while those occupying the first and third quadrants will remain in the same quadrants. This is a consequence of the momentum constraints specified in Table \ref{tab:mapping}.   For $(\omega,2\omega, \phi=\pi/2)$  and  $(\omega,2\omega, \phi=3\pi/4)$ fields, we expect the distributions to be L-shaped and mainly located along the positive $p_{n\parallel}=0$ half axis. This happens because there is only one dominant event per half cycle, and the most probable momentum of the second electron vanishes.  These predictions are summarized in Figs.~\ref{fig:diagrams}(d) and (e), respectively. For $(\omega,2\omega, \phi=\pi)$ [Fig.~\ref{fig:diagrams}(f)], $P_{1b}$ has become more important than $P_{1a}$ [see Fig.~\ref{fig:1ew2wpart}(f)] for the first electron, while the contributions of $O_{1a}$ and $O_{1b}$ are identical and centered at positive $p_{2\parallel}$. Putting these features together, one may construct a diagram in which the negative half axes are populated and occupy mainly the second and fourth quadrants of the parallel momentum plane, as shown in Figs.~\ref{fig:diagrams}(f). Note that in this case, the event $O_{1b}$ actually corresponds to the event $O_{2b}$ for $(\omega,2\omega, \phi=\pi/2,3\pi/4)$. Due to the fact that the original  $O_{1b}$ event is completely suppressed, the event $O_{2b}$ is named $O_{1b}$.
\subsection{Momentum constraints and distributions}
\begin{figure}[h]
    \centering
   \includegraphics[width=\linewidth]{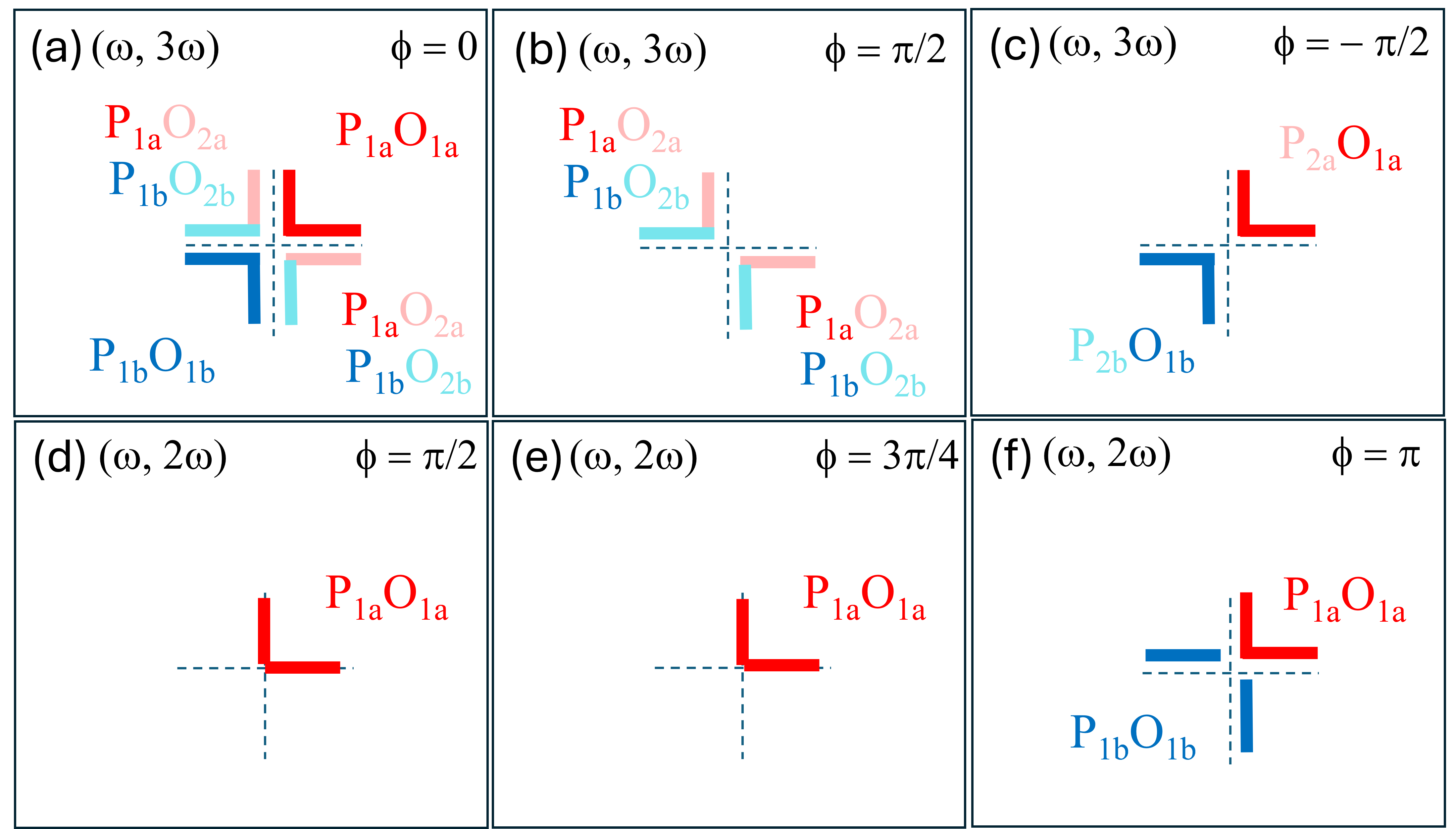}
    \caption{Diagrammatic representation of the dominant events for RESI with  $(\omega, 3\omega)$ and $(\omega, 2\omega)$ driving fields (upper and lower rows, respectively). Panels (a), (b), and (c) indicate the dominant events for an $(\omega, 3\omega)$ field with $\phi=0$, $\phi=\pi/2$ and $\phi=-\pi/2$, respectively, while panels (d), (e), and (f) outline the dominant events for an $(\omega, 2\omega)$ field with $\phi=\pi/2$, $\phi=3\pi/4$,  and $\phi=\pi$, respectively. The thick straight lines in the figure provide a schematic of what momentum regions of the $p_{1\parallel}p_{2\parallel}$ plane the correlated RESI distributions occupy. We consider the time intervals and events given in Table \ref{tab:mapping}.  Different colors indicate a time delay between events, while the same color indicates that one is dealing with the same event, with different regions associated with symmetrization. The colors associated with the events were matched to those in Fig.~\ref{fig:fields}, with red and pink used for the pairs and orbits $P_{ia}$, $O_{ia}$, and blue and cyan for the pairs and orbits $P_{ib}$, $O_{ib}$, respectively. The events and orbits are indicated by the labels $P_{ij}O_{kl}$ in the momentum regions they occupy, where $i$ and $k$ are integers, and $j$ and $l$ are $a$ or $b$. The thick lines representing an event $P_{ij}O_{kl}$ are colored according to the convention adopted for the orbits $O_{kl}$ rather than the total event.}
    \label{fig:diagrams}
\end{figure}
\begin{figure*}[!htbp]
\centering
\includegraphics[width=0.97\textwidth]{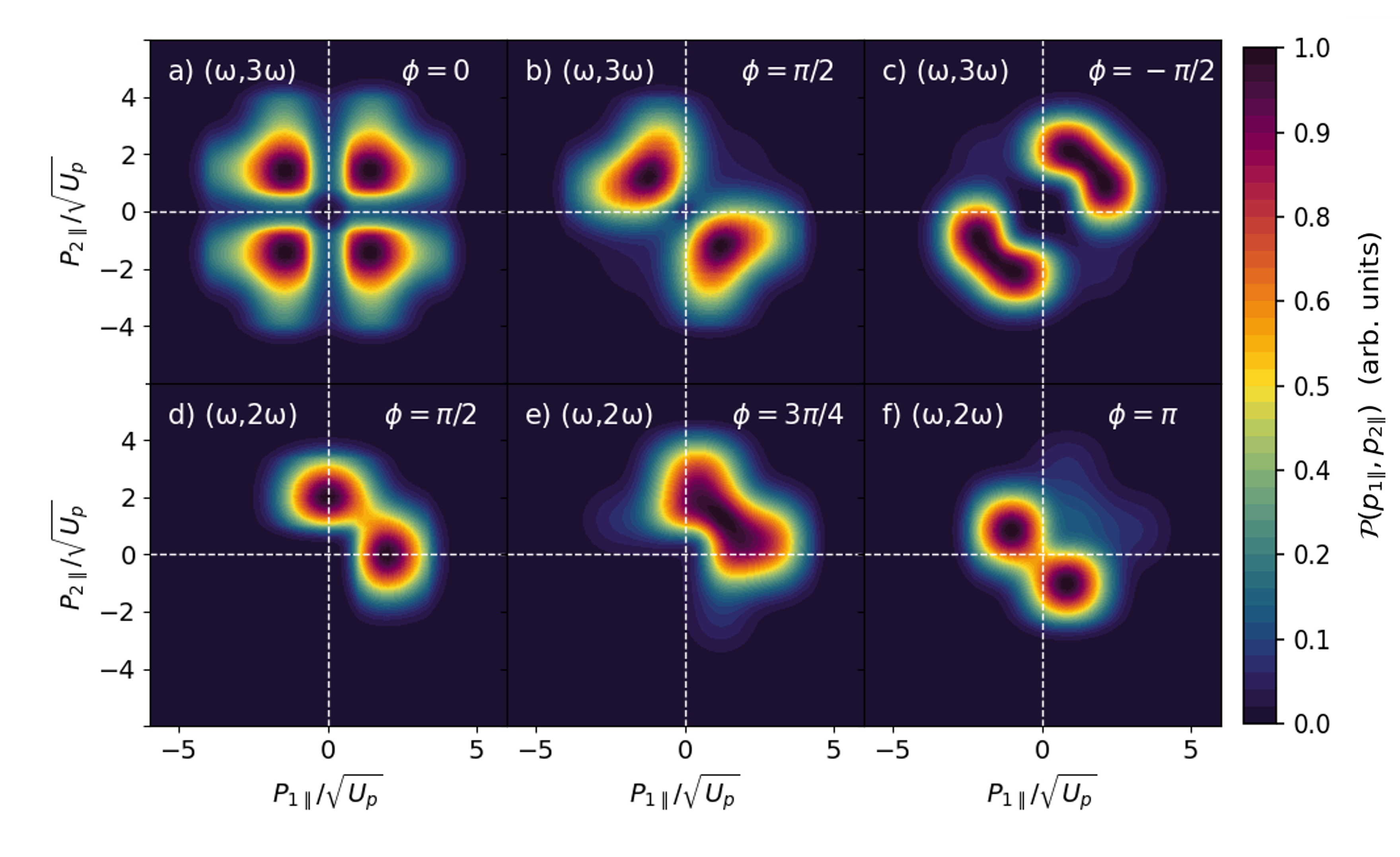}
    \caption{Incoherent momentum distributions without prefactors calculated for the ($\omega,3\omega$) (upper row) and ($\omega,2\omega$) (lower row) driving fields with the relative phase as indicated in the panels. Other driving-field parameters are the same as in the corresponding panels of Fig.~\ref{fig:fields}. The intensity of the $\omega$ field component and the fundamental wavelength are the same as in Fig.~\ref{fig:1ew3wpart}.}
    \label{fig:incoherent}
\end{figure*}

The two-electron RESI momentum distributions, plotted in Fig.~\ref{fig:incoherent}, agree with the above-mentioned predictions. Since we are assessing the dominance of specific events, we omit the prefactors and analyze the fully incoherent sum, which assumes that the contributions of different events as well as the contributions due to the symmetrization are taken into account incoherently. Including prefactors and quantum interference introduces additional biases and potentially breaks symmetries, and is detrimental to this assessment. Nonetheless, for the first electron, the contributions of the two solutions of a single pair are combined coherently using the uniform approximation as in \cite{Faria2002}. This is necessary because the artificial peaks present in the individual solutions' photoelectron yield alter the final distribution through integration via orthogonal momentum [see Fig.~3 in Ref.~\cite{Faria2003}].  This type of interference is washed out upon transverse momentum integration. 

For the ($\omega$, $3\omega$) field [see the upper row in Fig.~\ref{fig:incoherent}], all distributions possess the reflection symmetry about the diagonal or antidiagonal due to the half-cycle symmetry of the field. In addition, for relative phase $\phi=0$  [Fig.~\ref{fig:incoherent}(a)], the momentum distribution is symmetric with respect to the reflections $p_{1||}\rightarrow -p_{1||}$ and  $p_{2||}\rightarrow -p_{2||}$. Besides the equivalence of the contributions of $O_{1a,b}$ and $O_{2a,b}$ for the second electron, these additional symmetries require that the contributions of the $P_{1a,b}$ solutions of the first electron in the $p_{1||}>0$ and $p_{1||}<0$ parts of the momentum plane are equal as well. As shown in Figs.~\ref{fig:incoherent}(b) and (c), these additional reflection symmetries are not preserved for  $\phi=\pm \pi/2$. This is expected as, due to unequal field peaks, the relevance of the ionization events $O_{1a,b}$ and $O_{2a,b}$ is unequal. It causes the RESI distributions to occupy the first and third quadrants of the $p_{1\parallel}p_{2\parallel}$ plane for $\phi=-\pi/2$, and the second and fourth quadrants for $\phi=\pi/2$, as predicted in Fig.~\ref{fig:diagrams}.

The parallel momentum distributions obtained using the $(\omega$,$2\omega)$ field are displayed in the lower row of Fig.~\ref{fig:incoherent}. In this case, the reflection symmetry about the diagonal is always broken due to the absence of the half-cycle symmetry of the field.  The reflection symmetry about the antidiagonal is preserved. For relative phase $\phi=\pi/2$, plotted in Fig.~\ref{fig:incoherent}(d), the distributions occupy predominantly the positive $p_{n\parallel}$, $n=1,2$ half axes, according to the predictions in Fig.~\ref{fig:diagrams}(d). This is due to the $P_{1a}O_{1a}$ event being dominant. Furthermore, the probability densities are centered around the axes, which is due to $E(t_{1a})$ at the dominant ionization times for the second electron being reflection symmetric around its maxima. Some of these features, such as the positive momenta half-axis being populated, are preserved for $\phi=3\pi/4$ [see Fig.~\ref{fig:incoherent}(e) and the mapping in Fig.~\ref{fig:diagrams}(e)]. This is not surprising, as $P_{1a}O_{1a}$ still dominates. However, instead of being symmetric around $p_{n\parallel}=0$, $n=1,2$, the distributions are skewed towards the first quadrant of the parallel momentum plane. This is a consequence of the reflection symmetry around the field extrema being broken: although $p_{2\parallel}=0$ is still the most probable momentum with which the second electron will be freed, the field gradients differ for the positive and negative momentum regions in such a way that ionization is favored for $p_{2\parallel}>0$. Finally, for $\phi=\pi$, the dominant event is $P_{1b}O_{1b}$. The pair $P_{1b}$ populates mainly momentum regions for which $p_{1\parallel}<0$, and the events $O_{1a,b}$ are most probable for  $p_{2\parallel}>0$. Therefore, the distributions have moved to the second and fourth quadrants of the parallel momentum plane, in agreement with the mapping in Fig.~\ref{fig:diagrams}(f). Nonetheless, one sees a faint probability density in the first and third quadrants.  They stem from $P_{1a}O_{1a}$, which is expected to populate these regions. 
 
\subsection{Influence of prefactors}
\label{sec:prefactors}
\begin{figure*}[!htbp]
\centering
\includegraphics[width=0.97\textwidth]{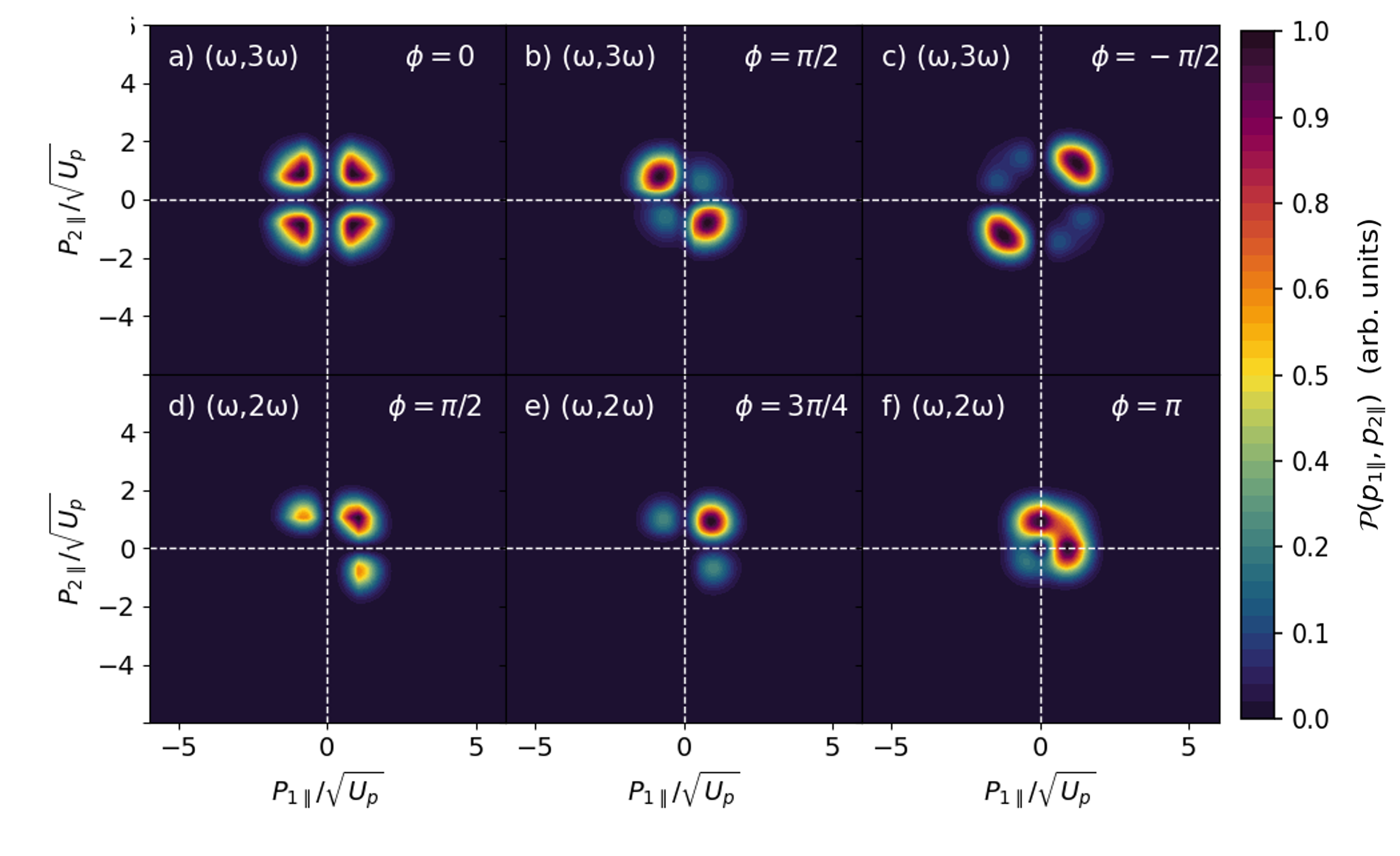}
    \caption{Incoherent momentum distributions with prefactors in the velocity gauge calculated using hydrogenic wavefunctions for the ($\omega$,$3\omega$) (upper row) and ($\omega$,$2\omega$) (lower row) driving fields with the relative phase as indicated in the panels. Other driving-field parameters are the same as in the corresponding panels of Fig.~\ref{fig:fields}. The intensity of the $\omega$ field component and the fundamental wavelength are the same as in Fig.~\ref{fig:1ew3wpart}.}
    \label{fig:hydrogenic}
\end{figure*}

After analyzing the momentum distributions calculated without prefactors, we now turn our attention to the influence of the prefactors on these distributions. Upon incorporation of the prefactors, all the distributions are narrowed and occupy a much smaller region of the momentum space. In particular, the ionization prefactor for the second electron, $V_{\mathbf{p}_{2e}}$, plays a critical role in determining the shapes of the momentum distributions, so it will be our main focus. In our previous publications, this prefactor led to practically identical results in the velocity and length gauge \cite{Shaaran2010}. However, in the present work, this is not necessarily the case.  A non-vanishing vector potential with complex arguments may lead to very different results in the length and velocity gauges, as well as counter-intuitive features. 

When $V_{\mathbf{p}_{2e}}$ is computed using the velocity gauge and hydrogenic wavefunctions [the expression can be found in \cite{Maxwell2015,Hashim2024}, while the shape and the mapping onto the $p_{1\parallel}p_{2\parallel}$ plane are presented in Fig. 8 in \cite{Hashim2024}], the distributions  [presented in Fig.~\ref{fig:hydrogenic}] exhibit a splitting along the axes characteristic for $p$ states [also see Fig.~8(g) in \cite{Hashim2024}]. This is particularly noticeable for the fields and phases in Fig.~\ref{fig:hydrogenic}(b)-(e). 
For example, in Figs.~\ref{fig:hydrogenic}(b) and (c), the prefactor splits the trailing edge of the distributions crossing $p_{n||}=0$, $n=1,2$, while in Figs.~\ref{fig:hydrogenic}(d) and (e) its effect is even more critical, as it suppresses the maxima in the positive half axes $p_{n\parallel}\geq 0$, $n=1,2$ [see Figs. \ref{fig:incoherent}(b) to (e) for comparison]. For the $(\omega,2\omega)$, $\phi=\pi$ case, the nodes along the axes are not present, and instead, the prefactor shifts the distribution to the axes. This is because, in the dominant $P_{1b}O_{1b}$ event, $O_{1b}$ is centered in the positive half $p_{2||}$ plane. It is thus affected mainly by the top lobe of the $p$-state prefactor, which causes $O_{1b}$ to become narrower and localized higher up in the positive half-plane. Upon combination with $P_{1b}$, the total contribution from this event moves from being very close to the origin in the negative-half plane to being centered around the axes.

Because $A(t)$ is no longer vanishing at the ionization time of the second electron, we can no longer neglect the vector potential in $V_{\mathbf{p}_{2e}}$ if it is computed in the length gauge. Within the SFA, particular care must be taken as the ionization prefactor exhibits a singularity, according to the saddle-point equation \eqref{eq:spati}, when the electron state is described by an exponentially decaying wave function, such as when using a hydrogenic basis.  For more details, see \cite{Faria2005}. This can be avoided by employing a Gaussian basis set to compute $V_{\mathbf{p}_{2e}}$, the expression for which is detailed in the part \ref{sec:gaussianprefactor} of the Appendix \ref{sec:appendix}, where we also verify that Gaussian and Hydrogenic basis sets both lead to the same momentum distributions in the velocity gauge.

\begin{figure*}[!htbp]
\centering
\includegraphics[width=0.97\textwidth]{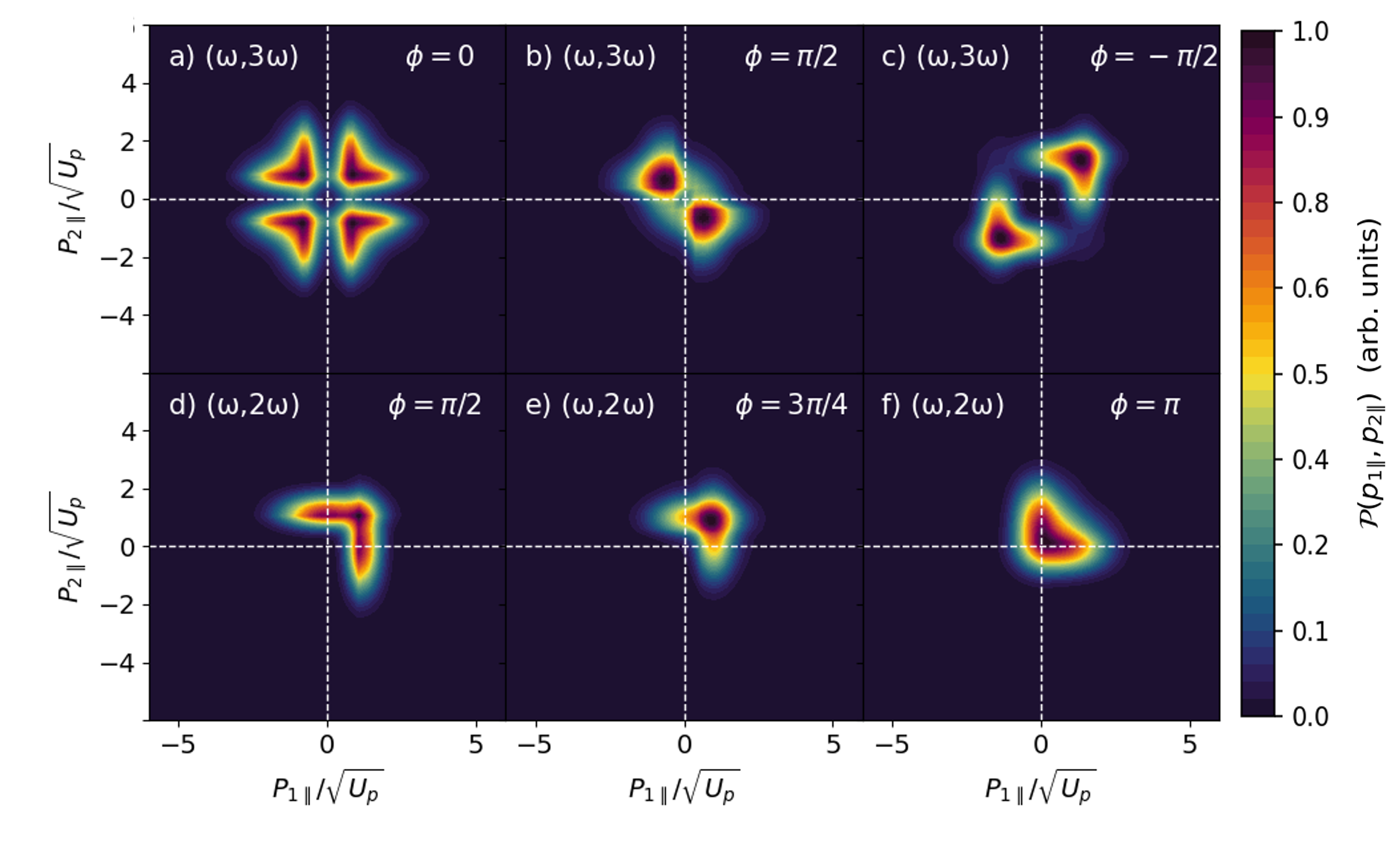}
    \caption{Incoherent momentum distributions with prefactors in the length gauge calculated using Gaussian wavefunctions and taking the complex second-electron ionization time for the ($\omega$,$3\omega$) (upper row) and ($\omega$,$2\omega$) (lower row) driving fields with the relative phase as indicated in the panels. Other driving-field parameters are the same as in the corresponding panels of Fig.~\ref{fig:fields}. The intensity of the $\omega$ field component and the fundamental wavelength are the same as in Fig.~\ref{fig:1ew3wpart}.}
    \label{fig:complexgaussian}
\end{figure*}

Next, we discuss the effect of the prefactor on the length-gauge RESI distribution [see Fig.~\ref{fig:complexgaussian}]. The shape of the excited state influences the prefactor through the spherical harmonic term in Eq.~\eqref{eq:Gaussian}, 
which can be expressed in terms of the momenta and the vector potential - for details, see \ref{sec:gaussianprefactor} of the Appendix \ref{sec:appendix}. 
Figure~\ref{fig:complexgaussian} shows a narrowing and slight sharpening of the distributions but the overall shape, notably the locations of the nodes and maxima are not significantly altered in comparison to the distributions without prefactors (cf. the corresponding distributions presented in Fig.~\ref{fig:complexgaussian} and Fig.~\ref{fig:incoherent}). These results are counterintuitive, as one would expect that a nonvanishing vector potential in the length-gauge prefactor would shift the maxima and/or the nodes of the prefactors away from the $p_{n\parallel}$ axes. To better understand this shape, a more detailed examination of the vector potential is necessary because, within the saddle-point framework, the vector potential of a linearly polarized bichromatic field is a sum of cosine functions with complex arguments, which can be written as a combination of ordinary and hyperbolic trigonometric functions - see \ref{sec:Aappendixlength} of the Appendix \ref{sec:appendix}. In Appendix  \ref{sec:Aappendixreal}, we analyze this effect in more depth by considering real arguments in the length-gauge prefactors. The exception is panel (f) associated with $(\omega,2\omega)$ field with the relative phase $\phi=\pi$, for which the prefactor is large enough to transfer the distribution from the negative half-plane to the origin. The impact of the prefactor for the $p$ states in the length gauge is much subtler than anticipated. Using similar reasoning, it can be shown that the impact of the prefactor for the $d$ states is of relatively little importance as its shape depends on the square of the argument $q_{2}$ of the spherical harmonic [for details see Eqs.~(\ref{eq:Gaussian})-(\ref{eq:thetaarg}) in the Appendix].

\begin{figure*}[!htbp]
\centering
\includegraphics[width=0.97\textwidth]{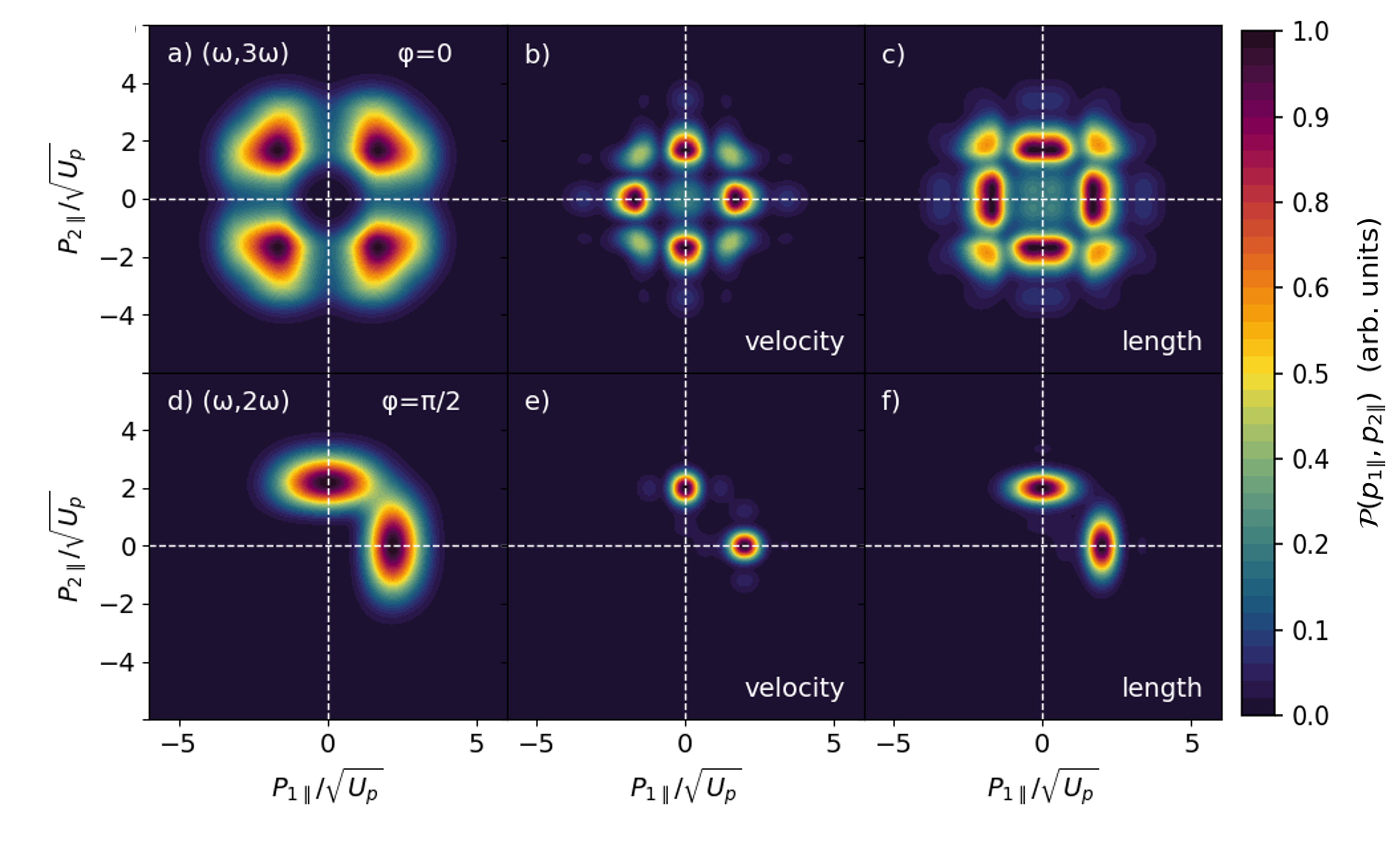}
   \caption{Incoherent momentum distributions without prefactors (first column), with hydrogenic prefactors in the velocity gauge (second column) and with gaussian prefactors in the length gauge (third column), calculated for the ($\omega$,$3\omega$), $\phi=0$ (upper row) and ($\omega$,$2\omega$), $\phi=\pi/2$ (lower row) driving fields. Other driving-field parameters are the same as in the associated panels of Fig.~\ref{fig:fields}. The intensity of the $\omega$ field component and the fundamental wavelength are the same as in Fig.~\ref{fig:1ew3wpart}.}
    \label{fig:3p4s}
\end{figure*}

For $s$ states, the overwhelming contribution comes from the radial integral, which is a confluent hypergeometric function. For real arguments, such as when $A(t)$ is excluded,  the function decays (grows) exponentially for large negative (positive) arguments. Adding the vector potential $A(t)$ shifts the real part of the argument in the prefactor, depending on the field symmetry. This shift localizes the prefactor near the peaks of the momentum distributions. Figure~\ref{fig:3p4s} shows this effect for the $3p \rightarrow 4s$ excitation channel, using two different fields: $(\omega, 3\omega)$ with $\phi=0$ and $(\omega, 2\omega)$ with $\phi=\pi/2$ (top and bottom rows, respectively). These parameters were chosen to illustrate cases where $A(t)$ is nonzero and zero, respectively. The first column [Figs.~\ref{fig:3p4s}(a) and (d)] shows the results without prefactors. As expected, they resemble those obtained for the $3s \rightarrow 3p$ excitation channel, but the more weakly bound $4s$ state leads to broader distributions than in Fig.\ref{fig:incoherent}. Including the velocity-gauge prefactor $V_{\mathbf{p}_2e}$ shifts the distributions to along the axes [Figs.~\ref{fig:3p4s}(b) and (e)], as is typical for $s$ states \cite{Shaaran2010}. This prefactor also introduces radial nodes, which appear as straight lines along the momentum axes [see Fig.~8(j) in \cite{Hashim2024}], causing the suppression in Fig.~\ref{fig:3p4s}(b) at non-zero parallel momenta. 
For the $(\omega, 2\omega)$, $\phi=\pi/2$ field, these nodal lines make the distribution narrower but still centered along $p_{n\parallel} = 0$ [Fig.\ref{fig:3p4s}(e)]. A similar effect is seen with the length-gauge prefactor [Figs.\ref{fig:3p4s}(c) and (f)], which also shifts the distribution along the axes and has a comparable nodal structure. However, the length-gauge distributions are more extended due to the imaginary part of the confluent hypergeometric function, which introduces oscillations when the argument is complex. This elongation is also observed for the $(\omega, 2\omega)$, $\phi=\pi/2$ field [compare Figs.~\ref{fig:3p4s}(e) and (f)].

\section{Conclusions}
\label{sec:conclusions}
In summary, we have studied the effect of the field symmetries and dominant events on the shapes of RESI photoelectron momentum distributions with two-color linearly polarized driving fields. We focused on how specific events populate the $p_{1\parallel}p_{2\parallel}$ plane and investigated the effect coming from the dynamics of each electron in detail. Furthermore, we have looked at the momentum bias associated with the geometry of the bound states, which, in our framework, is introduced by prefactors. In order to make an unambiguous assessment, we left two-electron quantum interference out.

A key difference between RESI in bichromatic fields and in the monochromatic fields \cite{Shaaran2010,Shaaran2010a,Maxwell2015,Maxwell2016} or few-cycle pulses \cite{Shaaran2012,Faria2012,Hashim2024} previously studied by us, is that, in general, it is not possible to relate a maximum of the field with a zero crossing of the vector potential, not even approximately. This comes from the approximate mapping $p_{2\parallel}=-A(t)$, and has major consequences for the ionization of the second electron and the resulting correlated RESI electron-momentum distributions.   For a monochromatic field and a few-cycle pulse, tunnel ionization for the second electron is most probable around a zero crossing of the vector potential, which means that the momentum transfer from the field to the electron is approximately vanishing. If there is no additional bias from a prefactor, this implies that the RESI distributions will be located around the momentum axes  $p_{n\parallel}=0$. 

In contrast, for a bichromatic field, if the intensity of the second wave is high enough, the vector potential corresponding to an extremum of $E(t)$ may be nonvanishing. This implies that the maxima of the RESI distributions will move away from the momentum axes. A striking example was obtained for a $(\omega,3\omega)$ field with $\phi=0$, for which there is a strong suppression around the $p_{n\parallel}=0$ axes. In this case, there is also more than one event contributing per half-cycle and the same symmetries as for the monochromatic field hold, leading to a fourfold symmetric distribution.  
The prominence of these events was influenced by the relative phase of the $(\omega,3\omega)$ field, but kept the distributions centered at non-vanishing momenta. One should note, however, that this is not always the case, as exemplified by the results obtained by the $(\omega,2\omega)$ field with $\phi=0$. In this case, although the half-cycle symmetry is broken, the distributions are centered around the positive half axis $p_{n\parallel}=0$, $p_{m\parallel}\geq0$, $n\neq m$.

Moving the distributions away from the $p_{n\parallel}=0$ axes also means that, in principle, depending on the frequency ratios and relative phase, we can confine electron momentum distributions to specific regions of the parallel momentum plane. In the present work, this has been achieved with $(\omega,3\omega)$ fields, for which distributions were confined to the first and third quadrant for $\phi=-\pi/2$, and to the second and fourth quadrant for $\phi=\pi/2$. This confinement may be even more extreme if we are dealing with scenarios for which the half-cycle symmetry is broken, such as the $(\omega,2\omega)$ field. For instance, if one chooses $\phi=3\pi/4$, the distributions are located almost entirely in the first quadrant, although in this case, the shifts stem from the unequal gradients around the electric field maximum. Confinement in a specific momentum region and comparable contributions from different events are important if one wishes to consider coherent superpositions of events and assess quantum-interference effects.  

In order to perform this confinement, one must determine what causes a specific event to be dominant. The present results suggest a hierarchy of parameters. For the first electron, the most important factor determining whether a pair of orbits is dominant is the tunneling probability around the field extrema associated with specific events. An example was provided for the $(\omega,3\omega,\phi=-\pi/2)$ field, for which a longer orbit pair led to the dominant contributions. The second most important parameter is the electron's excursion amplitude in the continuum. This is exemplified by the $(\omega,2\omega,\phi=\pi/2)$ field when comparing the contributions of the pairs $P_{1b}$ and $P_{2b}$. Finally, the classically allowed region is superseded by the other factors, as shown in the discussion of the results by the $(\omega,2\omega,\phi=\pi)$ field. For the second electron, the tunneling probability is extremely important, and, depending on the circumstances, tends to skew the dominance of the correlated electron momentum distributions.
Establishing this hierarchy has been attempted before using few-cycle pulses and allocating a dominance parameter \cite{Hashim2024}, but a bichromatic field seems to provide more control. Still, we anticipate that a single number to determine dominance, such as what was proposed in \cite{Hashim2024} will not be sufficient due to the strong momentum dependence of the partial distributions.

Furthermore, the RESI distributions are only fourfold symmetric if the three symmetries associated with the monochromatic fields are retained in a two-color scenario. This is achieved for an $(\omega,3\omega)$ field with $\phi=0$, but not with the other phases studied here. Nonetheless, retaining the half-cycle symmetry means that the RESI distributions are reflection symmetric about both diagonals $p_{1\parallel}=\pm p_{2\parallel}$, while if this symmetry is broken only $p_{1\parallel}=p_{2\parallel}$ holds. This intuitively makes sense because the processes displaced by half a cycle, which populate opposite sides of the anti-diagonal, should give the same contributions if this symmetry is retained. 

Distributions peaked at non-vanishing momenta $p_{2\parallel}$, together with the mapping $p_{2\parallel}=-A(t)$, means that extra care must be taken when incorporating prefactors in the electron-momentum distributions.  This will influence the ionization prefactor $V_{\mathbf{p}_2e}$ of the second electron considerably. Although we incorporated both rescattering and ionization prefactors, our discussions focus on $V_{\mathbf{p}_2e}$ as it is instrumental for determining the shapes of the RESI distributions. In particular, $A(t) \neq 0$ implies that calculating $V_{\mathbf{p}_2e}$ in the velocity and length gauges may lead to very distinct results, in contrast to what we observed for monochromatic fields \cite{Shaaran2010} and few-cycle pulses \cite{Hashim2024}. In order to avoid the bound-state singularities that appear in the length gauge, we employ a Gaussian basis to perform these calculations. The choice of gauge in the SFA has been the subject of long-standing debate \cite{Bauer2005,Ivanov2005,Smirnova2007b}. In our previous studies, a vanishingly small vector potential around the most probable ionizaton times meant that, in practice, this question could be put aside. However, the present work shows that, for bichromatic fields with waves of comparable strengths, care must be taken.  

Using the velocity gauge, the prefactor is centered about the origin. It causes a narrowing of the distributions, and the distributions take on the expected shape associated with the excited state: $p$ states lead to suppressions about $p_{n\parallel}$, $n=1,2$ axes and $s$ states mainly narrow the distributions.
Incorporating $A(t)$ has different effects for $p$ and $d$ states, compared to $s$ states. For $p$ and $d$ states, the shape is determined predominantly by the spherical harmonics, whilst, for $s$ states, the shape is determined by the confluent hypergeometric function. The present work shows that shifts in the prefactors predicted by the classical mapping $\mathbf{p}_2=-\mathbf{A}(\mathrm{Re}[t])$ do not hold and are counteracted by the imaginary part of this tunneling time. This is counterintuitive and shows limitations in the classical mappings, which behave as expected if the argument of the prefactors is forced to be real (see Appendix \ref{sec:Aappendixreal}). For discussions of the role of $\mathrm{Im}[t]$ in a broader, Coulomb-distorted context see \cite{Torlina2013,Maxwell2018b} for high-order harmonic generation and photoelectron holography, respectively.

Finally, in terms of using confinement in a specific momentum region to strengthen quantum interference in RESI, or even using orbit-based approaches beyond the SFA to model RESI distributions, a few issues must be taken into consideration. First, the mapping $p_{2\parallel}=-A(t)$, upon which many of the physical interpretations in this work rely, only holds if the long-range potential can be neglected in the continuum. Incorporating this potential is expected to influence the dynamics of the second electron substantially, and the mapping could fail if the acceleration caused by the potential in the continuum becomes significant. Therefore, since the shapes and maxima of the RESI momentum distributions are critically affected by the second electron, it is important to understand the parameter ranges for which this mapping is a good approximation, and when it is severely disrupted. Examples of this disruption have been provided in \cite{Zhang2014} for orthogonally polarized two-color fields.  Furthermore, the presence of the binding potential will lead to more orbits for the second electron, whose effects must be incorporated \cite{Rook2022,rodriguez2023}. Still, Coulomb-distorted two-color studies of photoelectron holography show that the symmetries investigated approximately using the SFA amplitude for direct ATI hold approximately when the Coulomb potential is incorporated \cite{Rook2022}. Fortunately, recent results indicate that, for the first electron, the dynamics of the rescattered orbits relevant to the present problem are well mimicked by the SFA \cite{Rook2024a}.
Second, the prefactor $V_{\mathbf{p}_2e}$ being dependent on $A(t)$, together with the times $t$ being complex in a saddle-point framework, mean that the momentum biases introduced may be counterintuitive with regard to classical symmetry arguments. For incoherent sums regarding symmetrization and events, we have verified in the present work that the effects are subtle in comparison to taking $A(t)=0$. Nonetheless, it is not clear how this extra dependence will influence coherent sums and the interference patterns in the correlated two-electron probability densities. Although, as shown in \cite{Maxwell2015,Maxwell2016,Hashim2024}, the RESI interference patterns are determined primarily by the phase differences stemming from the semiclassical action, time-dependent prefactors may cause some loss of contrast.  This could neutralize or reduce the increased overlap caused by confining the dominant contributions to RESI distributions to specific momentum regions. Answers to those open questions require further investigation.

\textbf{Acknowledgements: } Discussions with T. Rook and F. Mountford are gratefully acknowledged. This work has been funded by the UK Engineering and Physical Sciences Research Council (EPSRC) (Grant No. EP/T019530/1) and by UCL. D. Habibovi\'{c} thanks UCL for its kind hospitality. 

\appendix
\section{Prefactors}\label{sec:appendix}
\subsection{Prefactor using Gaussian basis}\label{sec:gaussianprefactor}
Using a linear combination of Gaussian-type orbitals (GTOs) to represent the radial part of the wavefunction, the ionization prefactor can be written as:
\begin{equation}
\begin{aligned}
V_{\mathbf{p}_2e}= & (-i)^{l_e} p_2^{l_e} 2^{(-1.5-l_e)} Y_{l_e}^0\left(\theta_{p_2}, \phi_{p_2}\right) \sum_{i=1}^{N} \alpha_i^{-1-l_e} c_{i}\\
& \frac{\Gamma(1+l_e)}{\Gamma(\frac{3}{2}+l_e)} { }_1F_1\left(1+l_e ; \frac{3}{2} + l_e ; -\frac{p_2^2}{4\alpha_i}\right),
\label{eq:Gaussian}
\end{aligned}
\end{equation}

where 
\begin{equation}
\begin{aligned}
\theta_{p_2}= & \cos^{-1}(q_{2}),
\label{eq:thetap2}
\end{aligned}
\end{equation}

\begin{equation}
\begin{aligned}
p_{2} = & \sqrt{[p_{2\parallel} + A(t)]^2 + p_{2\perp}^2},
\end{aligned}
\label{eq:p2}
\end{equation}
and
\begin{equation}
\begin{aligned}
q_{2} = & \frac{p_{2\parallel} + A(t)}{p_2},
\end{aligned}
\label{eq:thetaarg}
\end{equation}
$l_e$ is the orbital angular momentum of the excited state, and $c_i$ and $\alpha_i$ are the coefficients and exponents of the Gaussian basis, respectively. 

\begin{figure*}[!htbp]
\centering
\includegraphics[width=0.97\textwidth]{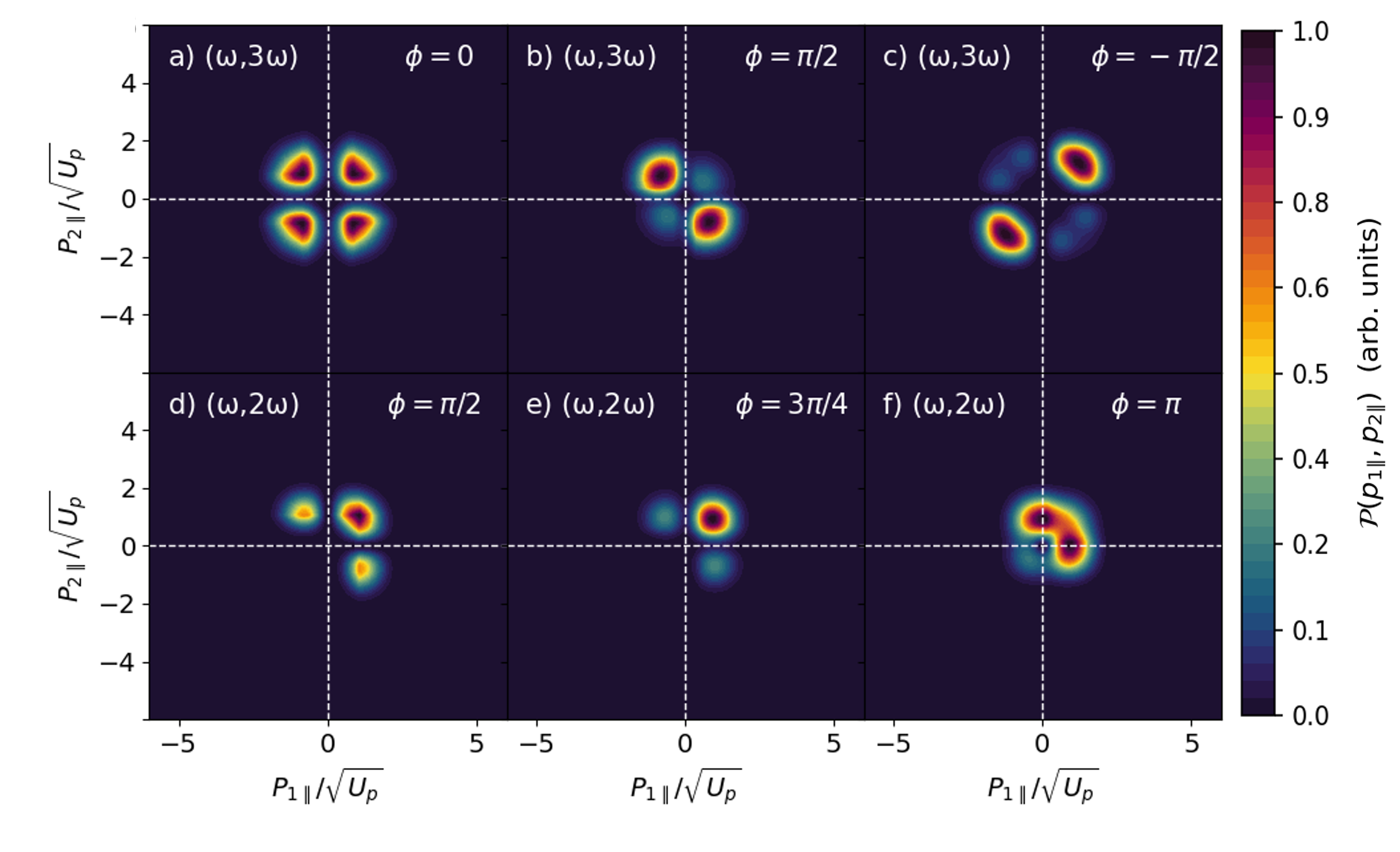}
    \caption{Incoherent momentum distributions with prefactors in the velocity gauge calculated using Gaussian wavefunctions and taking only the real part of the direct-electron ionization time for the ($\omega$,$3\omega$) (upper row) and ($\omega$,$2\omega$) (lower row) driving fields with the relative phase as indicated in the panels. Other driving-field parameters are  the same as in the corresponding panels of Fig.~\ref{fig:fields}. The intensity of the $\omega$ field component and the fundamental wavelength are the same as in Fig.~\ref{fig:1ew3wpart}.}
    \label{fig:gaussianappendix}
\end{figure*}

For the velocity gauge, we neglect the vector potential $A(t)$. This corresponds to assuming that the field-dressed momentum remains effectively unaltered by the external field. The incoherent momentum distributions computed using GTOs in the velocity gauge (presented in Fig.~\ref{fig:gaussianappendix}) demonstrate good qualitative agreement with those obtained using hydrogenic wavefunctions (Fig.~\ref{fig:hydrogenic}). 

\subsection{Two-electron momentum distributions with real-time prefactors}
\label{sec:Aappendixreal}

\begin{figure*}[!htbp]
\centering
\includegraphics[width=0.97\textwidth]{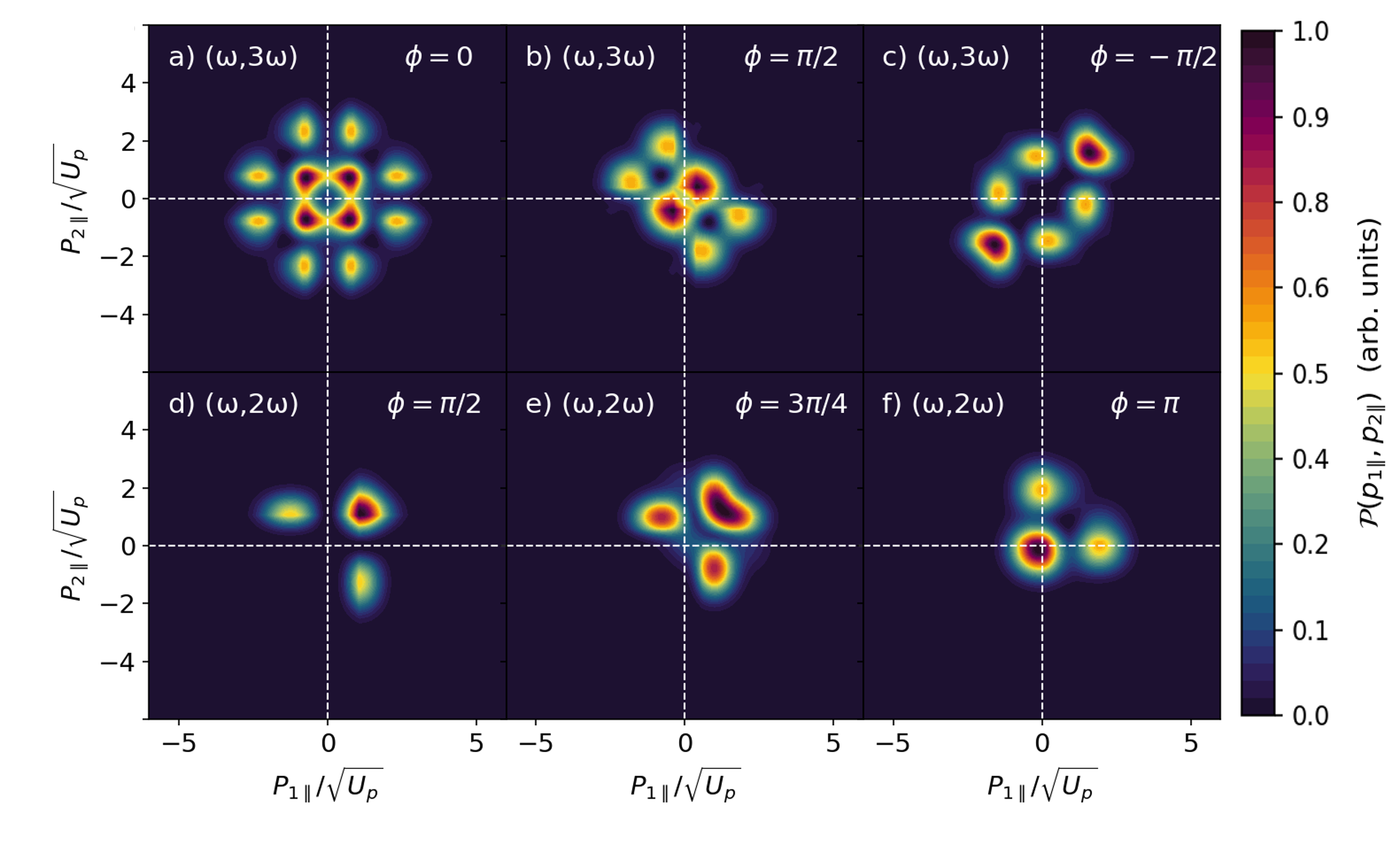}
    \caption{Incoherent momentum distributions with prefactors in the length gauge calculated using Gaussian wavefunctions and taking only the real part of the second-electron ionization time for the ($\omega$,$3\omega$) (upper row) and ($\omega$,$2\omega$) (lower row) driving fields with the relative phase as indicated in the panels. Other driving-field parameters are the same as in the corresponding panels of Fig.~\ref{fig:fields}. The intensity of the $\omega$ field component and the fundamental wavelength are the same as in Fig.~\ref{fig:1ew3wpart}.}
    \label{fig:realgaussian}
\end{figure*}

When only the real component of the ionization time $t$ is taken (Fig.~\ref{fig:realgaussian}), the shape of the vector potential remains as shown in Fig.~\ref{fig:fields}. The sum of the cosines merely causes a shift in the arguments of the spherical harmonics and hypergeometric functions. Therefore, the prefactor retains the same shape as when $A(t)$ vanishes, except that the distribution is now shifted along the $p_{2\parallel}$ axis. In other words, instead of being located at $p_{n\parallel}=0$, $(n=1,2)$, the nodes are now located at $p_{n\parallel}=\mp A(\mathrm{Re}[t])$, where $t$ is the ionization time for which $\mathrm{Im}[t]$ has a minimum. The signature of the excited $p$ state is perfectly aligned with the location of the momentum distributions, causing the splitting of the distributions associated with each event. This is particularly evident in the top row of Fig.~\ref{fig:realgaussian}. Figure \ref{fig:realgaussian}(a) shows that now the distribution in each quadrant has the characteristic $p$-state nodes. In Fig.~\ref{fig:realgaussian}(c), some distribution along the axes is retained because a node of the prefactor no longer occurs at the origin. For panel (d), the suppression still occurs at the axes due to $A(\mathrm{Re}[t])$ vanishing for $(\omega, 2\omega)$ field with the relative phase $\phi=\pi/2$. This behavior is also approximately seen in panel (e) [$(\omega, 2\omega)$, $\phi=3\pi/4$]. In panel (f), the prefactors cause the distribution to shift from the second and fourth quadrants to the axes and into the first quadrant. In addition, the $V^{(\mathcal{C})}_{\mathbf{p}_1e,\mathbf{k}g}$ prefactor causes a narrowing of the distribution and shifts it towards the origin. Although the $V^{(\mathcal{C})}_{\mathbf{p}_2e}$ prefactor causes a suppression at the axes, we have verified that the partial momentum distribution of the second electron is asymmetric around the origin and very steep close to the axis, thus when considered in conjunction with the bias introduced by the first electron, it is washed out. 

\subsection{Length-gauge prefactor using complex time}\label{sec:Aappendixlength}

Within the framework of the saddle-point method, the time $t$ which appears in the vector potential $A(t)$ defined by Eq.~\eqref{eq:Afield} is complex, i.e., $t = \mathrm{Re}[\omega t] + i \mathrm{Im}[\omega t]$. In this case, each of the cosine functions can be rewritten as

\begin{eqnarray}
\cos(\omega t) &=& \cos(\mathrm{Re}[\omega t]) \cosh(\mathrm{Im}[\omega t])\nonumber \\&&- i \sin(\mathrm{Re}[\omega t]) \sinh(\mathrm{Im}[\omega t]).
\end{eqnarray}

In Figs.~\ref{fig:2ew2wpart} and \ref{fig:2ew3wpart}, we have presented the results for $\mathrm{Im}[\omega t]$ for $p_{2\perp}=0$. We have also verified that $\mathrm{Im}[\omega t]>0$ for all other values of $p_{2\perp}$ and vanishing $p_{2\parallel}$. More specifically, a non-vanishing $p_{2\perp}$ merely shifts the times upwards. In addition, we have checked that $\mathrm{Re}[\omega t]>0$, for all fields and events used in this study and that $\mathrm{Re}[\omega t]>\mathrm{Im}[\omega t]$ in all cases. With a complex time, the behavior of $A(t)$ in the complex plane is determined by the interplay of the ordinary and hyperbolic trigonometric functions subject to these constraints. 

Now we consider what happens to $A(t)$, and hence $q_{2}$, for different values of $\mathrm{Re}[\omega t]$ and $\mathrm{Im}[\omega t]$ to understand the behavior of $q_{2}$, and thus how it affects the momentum distributions. Regardless of the time, at the origin i.e., at $(p_{2\parallel}, p_{2\perp}) = (0,0)$, $q_{2} = 1$.

For small real and imaginary parts of the time, Taylor-series approximations, $\cos(\mathrm{Re}[\omega t]) \approx 1$,  $\sin(\mathrm{Re}[\omega t]) \approx \mathrm{Re}[\omega t]$,  $\cosh(\mathrm{Im}[\omega t]) \approx 1$, and $\sinh(\mathrm{Im}[\omega t]) \approx \mathrm{Im}[\omega t]$ for the trigonometric functions can be utilized. This results in $A(t)$ having a constant positive real part, and a negative imaginary part directly proportional to $\mathrm{Re}[\omega t] \cdot \mathrm{Im}[\omega t]$. If either of $\mathrm{Re}[\omega t]$ or $\mathrm{Im}[\omega t]$ is sufficiently small, the imaginary component of $A(t)$ may be neglected. In this case, the sign of $q_{2}$  will be positive for
\begin{equation}
\begin{aligned}
p_{2\parallel} > \frac{2 \sqrt{U_p}}{\sqrt{ \frac{1}{r^2} + \frac{\xi^2}{s^2} }} \left(1 + \frac{\xi}{s}\right)
\end{aligned}
\end{equation}
and negative elsewhere.  

When $p_{2\parallel}$ is negligible, but $p_{2\perp}$ is large, $q_{2}$ can be approximated by
\begin{equation}
\begin{aligned}
q_{2} = & \frac{A(t)}{\sqrt{A^2(t) + p_{2\perp}^2}}.
\end{aligned}
\end{equation}
In this limit, $q_{2}$ approaches 1 for vanishing $p_{2\perp}$ and a given ionization time. In addition, it can be verified that  $ \operatorname{Im}[\omega t]$ increases as $p_{2\perp}^2$ increases so that, for a given $p_{2\perp}$, the imaginary component of $A(t)$ increases. For large enough values of $p_{2\perp}$, the function may approach 1/$p_{2\perp}$. 
Conversely, small $p_{2\perp}$ and large $p_{2\parallel}$ results in the shape determined by
\begin{equation}
\begin{aligned}
q_{2} = & \frac{p_{2\parallel} + A(t)}{\sqrt{(p_{2\parallel} + A(t))^2}}.
\end{aligned}
\end{equation}
In this limit, the function approaches $\pm 1$ depending on the magnitude of $p_{2\parallel}$ regardless of the individual real and imaginary components. The term $q_{2}$ is dependent on $p_{2\parallel}$ as well as on $p_{2\perp}^2$. When plotted in the $p_{2\parallel} p_{2\perp}$ plane, it exhibits regions parallel to the $p_{2\parallel}$ axis of varying magnitude. 

On the other hand, for large real and imaginary parts of the time, the ordinary trigonometric functions oscillate while the hyperbolic trigonometric functions both grow exponentially (since $\mathrm{Im}[\omega t]>0$). Eventually, the hyperbolic functions dominate, causing $A(t)$ to approach infinity. As a result, $A(t)$ dominates, and $q_{2} \rightarrow 1$ in this limit, regardless of the value of the momenta.

Finally, away from the large and small time limits, i.e., with real and imaginary parts of the time that are neither very small nor large, the oscillatory behaviour of $A(t)$ is retained. For $\mathrm{Re}[\omega t] >> \mathrm{Im}[ \omega t]$ the oscillatory components dominate. The amplitude of the oscillation is constrained by $p_{2 \perp}$ for negligible $p_{2  \parallel}$. For large $p_{2 \parallel}$ and negligible $p_{2 \perp}$, the oscillations are limited and stabilized around $p_{2\parallel}$. When the real part of the time is not significantly greater than the imaginary part, we obtain finite complex values for $A(t)$ constrained by momenta. 

Taking everything into account, we conclude that for all cases, the shape of the absolute value of the prefactor can be interpreted as fringes parallel to the axes, either constant or oscillating in value and symmetric about the axes.

\end{document}